\documentclass[a4paper,twocolumn, iop, twocolappendix, numberedappendix]{oja}
\usepackage{graphicx} 
\usepackage[T1]{fontenc}
\usepackage[utf8]{inputenc}
\usepackage{natbib}
\usepackage{fontawesome5}
\usepackage{multirow}
\usepackage{comment}
\usepackage{graphicx}	
\usepackage{amssymb}	
\usepackage{xspace}        
\usepackage[dvipsnames]{xcolor}
\usepackage{booktabs}
\usepackage{fancyhdr}
\usepackage{upgreek}
\usepackage{tensor}
\usepackage{times}
\usepackage{tikz}
\usetikzlibrary{angles, quotes, 3d}
\usepackage{newtxtext,newtxmath}

\usepackage[colorlinks,linkcolor=blue,citecolor=blue,urlcolor=blue ]{hyperref}
\usepackage{xurl} 

\newcommand{\review}[1]{\textcolor{black}{#1}}

\usepackage{cleveref}
\begin{document}
\title{An analytical model for the dispersion measure of fast radio burst host galaxies}
\author{Robert Reischke$^{\dagger}$, Michael Kova\v{c} and  Andrina Nicola}
\affiliation{Argelander-Institut für Astronomie, Universität Bonn, Auf dem Hügel 71, D-53121 Bonn, Germany}
\thanks{$^\dagger$\href{mailto:reischke@posteo.net}{reischke@posteo.net}\\ \phantom{c$\!\mkern-1.5mu$} \href{mailto:rreischke@astro.uni-bonn.de}{rreischke@astro.uni-bonn.de}}

 \author{Steffen Hagstotz}
\affiliation{Universitäts-Sternwarte, Fakultät für Physik, Ludwig-Maximilians Universität München, 
Scheinerstraße 1, D-81679 München, Germany and\\
Excellence Cluster ORIGINS, Boltzmannstraße 2, D-85748 Garching, Germany}

\author{Aurel Schneider}
\affiliation{Department of Astrophysics, University of Zurich, Winterthurerstrasse 190, 8057 Zurich, Switzerland}

\begin{abstract}
The dispersion measure (DM) of fast radio bursts (FRBs) is sensitive to the electron distribution in the Universe, making it a promising probe of cosmology and astrophysical processes such as baryonic feedback. However, cosmological analyses of FRBs require knowledge of the contribution to the observed DM coming from the FRB host. The size and distribution of this contribution is still uncertain, thus significantly limiting current cosmological FRB analyses.

In this study, we extend the baryonification (BCM) approach to derive a physically-motivated, analytic model for predicting the host contribution to FRB DMs. By focusing on the statistical properties of FRB host DMs, we find that our simple model is able to reproduce the probability distribution function (PDF) of host halo DMs measured from the CAMELS suite of hydrodynamic simulations, as well as their mass- and redshift dependence. Furthermore, we demonstrate that our model allows for self-consistent predictions of the host DM PDF and the matter power spectrum suppression due to baryonic effects, as observed in these simulations, making it promising for modelling host-DM-related systematics in FRB analyses. In general, we find that the shape of the host DM PDF is determined by the interplay between the FRB and gas distributions in halos. Our findings indicate that more compact FRB profiles require shallower gas profiles (and vice versa) in order to match the observed DM distributions in hydrodynamic simulations. Furthermore, the analytic model presented here shows that the shape of the host DM PDF is highly sensitive to the parameters of the BCM. This suggests that this observable could be used as an interesting test bed for baryonic processes, complementing other probes due to its sensitivity to feedback on galactic scales. We further discuss the main limitations of our analysis, and point out potential avenues for future work.
\end{abstract}

\maketitle
\section{Introduction}
Fast radio bursts (FRBs) are short transients lasting only a few milliseconds, with a frequency range from $\sim 100$ MHz to several GHz. 
The initial pulse is dispersed due to free electrons in the ionised galactic and intergalactic medium, resulting in a delayed arrival time of the pulse frequencies, $\Delta t(\nu) \propto \nu^{-2}$. The proportionality constant, known as the dispersion measure (DM), is related to the total column density of electrons along the line-of-sight to the FRB \citep[e.g.][]{2007Sci...318..777L,thornton_population_2013, petroff_real-time_2015, connor_non-cosmological_2016, champion_five_2016,chatterjee_direct_2017}.

Although the mechanism behind the radio emission is still debated, the isotropic occurrence and high observed DM of these bursts suggest an extragalactic origin for the majority of events \citep[though some might be Galactic, see][]{andersen_bright_2020}. Consequently, the DM can be used to study the distribution of diffuse electrons in the large-scale structure (LSS). Several researchers have proposed using DMs inferred from FRBs as a cosmological probe, either through the average dispersion measure up to a given redshift \citep{zhou_fast_2014,walters_future_2018,hagstotz_new_2022,macquart_census_2020,wu_8_2022,james_measurement_2022,reischke_consistent_2023-1,reischke_cosmological_2023,khrykin_flimflam_2024} or through the statistics of DM fluctuations \citep{masui_dispersion_2015,shirasaki_large-scale_2017,rafiei-ravandi_characterizing_2020,reischke_probing_2021,bhattacharya_fast_2020,takahashi_statistical_2021,rafiei-ravandi_chimefrb_2021,2022MNRAS.512.1730S,reischke_consistent_2022,reischke_calibrating_2023,2025OJAp....8E..72N}. While the former requires identifying the host galaxy to obtain an independent spectroscopic redshift estimate, the latter can be performed without it, as the mean of the dispersion can serve as a (noisy) estimate for the redshift.

Despite their potential, FRB-based cosmological probes face several challenges: First, localising FRBs to their host galaxies has so far only been possible for a relatively small sample, thus limiting the available data for cosmological applications. Second, the observed DMs receive contributions from the FRB host galaxy, the intergalactic medium (IGM), and the Milky Way, whose size is subject to significant systematic uncertainties. While models exist for the latter two contributions, the host contribution lacks a well-defined description. This becomes in particular impactful when considering budgeting of the DM for different sightlines \citep[see e.g.][]{niu_repeating_2022,connor_deep_2023,connor_gas_2024}

Recently, it has been shown that the host contribution as measured from hydrodynamic simulations follows a log-normal distribution \citep{Jaroszy_DM_2020,walker_dispersion_2024,medlock_probing_2024,theis_galaxy_2024}; see also \citep{mcquinn_locating_2014, 2020MNRAS.496L.106K} for previous simple analytical models. Additionally, the statistical properties of the host DM were shown to depend rather strongly on baryonic effects, such as feedback due to Active Galactic Nuclei (AGN) and Supernovae (SNe).
These feedback processes also affect the matter distribution in the Universe and therefore cosmological observables such as weak gravitational lensing (WL) \citep{2008ApJ...672...19R,van_daalen_effects_2012,semboloni_effect_2013,harnois-deraps_baryons_2015,huang_modelling_2019,chisari_impact_2018,chisari_modelling_2019, nicola_breaking_2022, reischke_calibrating_2023}. Currently, the amplitude and shape of these effects are still largely unknown, with hydrodynamic simulations yielding different results depending on the sub-grid feedback prescriptions employed. These systematic uncertainties severely limit the science return of current Stage-III experiments and will continue to do so for Stage-IV surveys, such as Euclid\footnote{\href{https://www.euclid-ec.org/}{https://www.euclid-ec.org/}}, the Rubin Observatory Legacy Survey of Space and Time (LSST)\footnote{\href{https://www.lsst.org/}{https://www.lsst.org/}} and the Roman space telescope\footnote{\href{https://roman.gsfc.nasa.gov/}{https://roman.gsfc.nasa.gov/}}, which obtain the largest fraction of their cosmological signal from highly non-linear scales, strongly affected by baryonic feedback.

While running a large number of hydrodynamic simulations with cosmologically-relevant box sizes in order to understand baryonic feedback is currently computationally prohibitive, several alternatives have been developed to model the effects of baryons on the matter distribution, for example halo model and emulator approaches \citep{mead_hydrodynamical_2020,mccarthy_improving_2020,angulo_bacco_2021,troster_joint_2022}. 
In \cite{schneider_new_2015,schneider_quantifying_2019,giri_emulation_2021} it has been shown that the effects of baryons can be included in gravity-only $N$-body simulations by displacing particles in halos. This procedure is called baryonification (or baryonic correction model, BCM), and is based on observationally-motivated, analytic matter, gas, and stellar profiles. The model has been shown to accurately reproduce matter power spectra obtained from hydrodynamic simulations for a range of feedback models, and has already been employed in several WL analyses \citep[e.g.][]{fluri_cosmological_2019,2022PhRvD.105h3518F,2022MNRAS.514.3802S,2024MNRAS.534..655B}.

In this work, we use the BCM to construct a physically motivated model for the host halo contribution to FRB DMs. While the parameters of this model should, in principle, be the same as those describing the impact of baryonic physics on the matter power spectrum, one should keep in mind that the shape of the matter power spectrum on scales accessible to LSS surveys is dominated by massive halos with $M> 10^{12}h^{-1}M_\odot$. In this work, we are concerned with a different regime, as FRB hosts are typically Milky-Way sized galaxies with masses $M\sim 10^{11}h^{-1}M_\odot$ or smaller. It is not clear, at this stage, how strongly the parameters of the BCM will vary with halo mass. The motivation of this paper is therefore not to provide constraints from low-mass halos as a lever for baryonic feedback in massive halos, but rather to develop an analytic, flexible model of the DM host contribution, which can eventually be used in a more general framework to link baryonic effects from galactic to cosmological scales. 

The paper is organised as follows: In Section \ref{sec:FRB_basics} we discuss the basics of FRBs and the different DM components. Section \ref{sec:host} introduces the BCM for the host DM. We compare and match our results to hydrodynamic simulations in Section \ref{sec:sim}. In Section \ref{sec:results}, we discuss the general properties of the model, including redshift, mass, and parameter dependence. Lastly, we conclude {in} Section \ref{sec:conclusion}. In all of this work, we fix the background cosmology to the values used in the CAMELS simulation's cosmic variance runs \citep{villaescusa-navarro_camels_2021,theis_galaxy_2024}, that is $\Omega_\mathrm{b}=0.049$, $h = 0.6711 $, $n_\mathrm{s} =0.9624$, $\sigma_8 = 0.8$, $\Omega_\mathrm{m}=0.3$, no massive neutrinos, no spatial curvature and a cosmological constant.

\section{Fast Radio Bursts basics}
\label{sec:FRB_basics}
A radio pulse emitted by an FRB source at position $\boldsymbol{x}$ and redshift $z$ is dispersed over a time $\Delta t$ via the following relation,
\begin{equation}
\label{eq:DM_def}
    \Delta t = \mathrm{DM}_\mathrm{obs}(\mathbf{x},z) \nu^{-2}\,,
\end{equation}
where $\nu$ is the frequency. This defines the observed DM, $\mathrm{DM}_\mathrm{obs}$, due to all free electrons along the line-of-sight. Quite generally, one can split up the electrons as residing in the Milky Way (MW), the LSS or the host galaxy of the FRB. In the literature, one sometimes further attempts to separate the contribution from the interstellar medium and the halo for the MW as well as the host \citep[see][for a most recent example]{khrykin_flimflam_2024}. The DM is given by the line-of-sight integral over the comoving electron density, $n_\mathrm{e}$:
\begin{equation}
\label{eq:dm_los}
    \mathrm{DM}_\mathrm{obs}(\mathbf{x},z) = \int_0^z \frac{n_\mathrm{e}(l)}{1+z(l)}\mathrm{d}l\,,
\end{equation}
where the factor $a=1/(1+z)$ in the integrand is used for the transformation into physical coordinates.
 Due to this structure, $\mathrm{DM}_\mathrm{obs}$ splits up into different contributions as well:
\begin{equation}
\label{eq:parts}
    \mathrm{DM}_\mathrm{obs}(\mathbf{x},z) = \mathrm{DM}_{\mathrm{host}}(z) + \mathrm{DM}_{\mathrm{MW}}(\mathbf{x}) + \mathrm{DM}_{\mathrm{LSS}}(z,\mathbf{x})\,,
\end{equation}
where we did not further split the MW and the host component, as this is not necessary in a halo model context. Note that the dependence on redshift, $z$, and position was made explicit in each component. The MW has no redshift dependence, and the host contribution does not depend on where the host galaxy is located. While the contribution of an individual host galaxy depends on the position of the FRB within the host, identifying the location of an FRB within a galaxy is observationally a very challenging task. With current FRB surveys, we can hence only make statistical predictions about the host properties, i.e. averaged over all possible positions of the FRB. Hence, the host contribution depends only explicitly on redshift.

\section{Host contribution model}\label{sec:host}
\subsection{Baryonification}

The BCM, originally proposed by \cite{schneider_new_2015,schneider_quantifying_2019} provides a physically-motivated and computationally efficient way to account for baryonic effects in $N$-body simulations by displacing particles in halos according to analytic density profiles. A large advantage of the BCM is that it is integrated trivially into halo model approaches of the large-scale structure, thus potentially providing a flexible model for the analysis of the total matter, but also the baryon distribution in the Universe. 
 Here we summarise the ingredients of the BCM and their associated analytic profiles.

In the BCM, the dark-matter-only (dmo) halo profiles are mapped into dark-matter-baryon (dmb) profiles by including the effects of dark matter, gas and stars \citep{schneider_new_2015, schneider_quantifying_2019,giri_emulation_2021}, according to  
\begin{equation}
\label{eq:baryonification}
    \rho_\mathrm{dmo}(r) \mapsto\rho_\mathrm{dmb}(r) = \rho_\mathrm{clm}(r) + \rho_\mathrm{gas}(r) + \rho_\mathrm{cen}(r)\,,
\end{equation}
with the profiles of collisionless matter (clm), gas, and the central galaxy (cen) respectively. The dmo profile is modelled as a truncated NFW profile \citep{navarro_universal_1997,baltz_analytic_2009}, and the clm profile contains, besides dark matter, also satellite galaxies as well as intra cluster stars. The analytic form of these profiles is observationally-motivated, and their particular shape is determined by the parameters of the BCM.

For a given set of model parameters, one can then perturb the halo profiles in pure gravity $N$-body simulations to obtain new profiles, accounting for the presence and feedback due to baryons. 

\begin{table}
 \renewcommand{\arraystretch}{1.4}
    \centering
    \begin{tabular}{cccc}\hline\hline
    Parameter & {SIMBA} & TNG & Description \\ \hline
   $\log_{10}$ $M_\mathrm{c}$ & $ {12.7}\pm 0.4 $ & $ 14.1 \pm 0.3$& mass scale gas \\
$\theta_\mathrm{ej}$ & ${7.95}\pm1.5 $ & $5.0 \pm 0.9$ &ejection radius \\
$\mu$ & ${0.19}\pm 0.11$ & $0.49\pm 0.18$ &slope of mass dependence \\
$\delta$ &${9.0}\pm 1.1$ & $10.1\pm 0.9$ & transitional slope \\
$\gamma$ &${1.0}\pm 0.3 $&$ 1.0\pm 0.4$ &outer profile slope \\
$\theta_\mathrm{co}$ &${0.017}\pm0.050$ & $0.009\pm 0.070$ & core radius \\
$\alpha$ &{2} & 2&stellar exponential slope  \\
$r_\mathrm{cut}[r_\mathrm{vir}]$ &${0.27}\pm0.13 $&$ 0.53\pm 0.10$& stellar cut-off radius
    \end{tabular}
    \caption{{Best fitting parameters and their Gaussian $1\sigma$ error used for the BCM to reproduce the host PDF measured in SIMBA and IllustrisTNG \citep{theis_galaxy_2024}. Throughout the paper, we will use the TNG values (second column) as fiducial. The corresponding PDFs can be seen in \Cref{fig:pdf_dm_SIMBA}.}}
    \label{tab:parameters}
\end{table}

In the BCM, the gas profile is modelled as a cored double-power law profile:
\begin{equation}
\label{eq:rhogas}
    \rho_\mathrm{gas}(r) \propto \frac{\Omega_\mathrm{b}/\Omega_\mathrm{m} - f^\star(M)}{\bigg[1+\left(\frac{r}{\theta_\mathrm{co}r_\mathrm{vir}}\right)\bigg]^{\beta(M)}\bigg[1+\left(\frac{r}{r_\mathrm{vir}\theta_\mathrm{ej}}\right)^\gamma\bigg]^{\frac{\delta - \beta (M)}{\gamma}}}\;,
\end{equation}
where $\theta_\mathrm{co}$ and $\theta_\mathrm{ej}$ parametrise the core and ejection radii, $\delta$ controls the outer slope, and the inner slope is determined by $\beta(M)$. The transition regime is described by $\gamma$.
In line with observations, $\beta$ is mass-dependent and parametrised as
\begin{equation}
\label{eq:beta}
    \beta(M_\mathrm{c},\mu) \coloneqq \frac{3(M/M_\mathrm{c})^\mu}{1 + (M/M_\mathrm{c})^\mu}\,.
\end{equation}
This allows for the gas profile to become shallower than the dmo NFW profile for $M < M_\mathrm{c}$, while being bounded from above as $\beta \leq 3$. \review{The parameters $M_\mathrm{c}$ and $\mu$ describe at which mass this transition happens and how quickly, respectively.}
In this work, halo masses are defined as the mass enclosed within a sphere of 200 times the critical density, and we further use the concentration-mass relation from \cite{Dutton_cold_2014}.

The gas profile is normalised according to the gas fraction, which depends on the matter and baryon densities in the Universe, $\Omega_\mathrm{m}$, $\Omega_\mathrm{b}$, and the total fraction of stars in halos, $f^\star$. Both the total stellar fraction and the fraction of stars in the central galaxy are given by a double power law:
\begin{equation}
\label{eq:star fraction}
    f^{\star/\mathrm{cen}} = 0.055\left[\left(\frac{M}{M_\mathrm{s}}\right)^{-\eta} + \left(\frac{M}{M_\mathrm{s}}\right)^{\eta_{\star/\mathrm{cen}}}\right]^{-1}\,,
\end{equation}
where $\eta=1.3$, $M_\mathrm{s} = 2.5\times 10^{11}h^{-1}M_\odot$ and the slopes are controlled by $\eta_{\star}$, $\eta_{\mathrm{cen}}$ respectively. The satellite stellar fraction is then simply obtained by subtracting the central fraction from the total fraction. This is different to the standard BCM model \citep{schneider_quantifying_2019}, where the stellar fraction is described by a \review{single power law which is a simplified version of the fit following abundance matching \citep{Moster_2012} as it only corresponds to halos with mass $M>10^{12}h^{-1}M_\odot$. Since we are considering lower halo masses in this work, the latter model can lead to unphysically large stellar fractions.}

The collisionless matter will contract and expand in response to the gravitational effect of the gas and central galaxy components, an effect called adiabatic relaxation. This can be captured through the relation
\begin{equation}
    \zeta = \frac{r_\mathrm{f}}{r_\mathrm{i}}-1=a_\zeta\left[\left(\frac{M_\mathrm{i}}{M_\mathrm{f}}\right)^{n_\zeta}-1\right]\,,
\end{equation}
where $r_\mathrm{i}$ and $r_\mathrm{f}$ are the initial and final radii of collisionless matter shells and $M_\mathrm{i}$ and $M_\mathrm{f}$ are the initial and final enclosed masses. Following \citet{2022MNRAS.512.1730S,2011MNRAS.414..195T}, we set $a_\zeta=0.3$ and $n_\zeta=2$. With this, we can write the clm distribution as a scaled truncated NFW profile \citep{navarro_universal_1997,baltz_analytic_2009}:
\begin{equation}
    \rho_\mathrm{clm} (r) = \left[\frac{\Omega_\mathrm{dm}}{\Omega_\mathrm{m}}+f^{\star}_\mathrm{sat}(M)\right]\frac{\rho_\mathrm{NFW}(r/\zeta)}{\zeta^3}\,,
\end{equation}
with $f^{\star}_\mathrm{sat}(M)$ obtained from $f^{\star}_\mathrm{cen}(M)$ and $f^\star(M)$ as discussed before.

For halos of group and cluster size, the stellar density is well represented by a low-concentration NFW profile. However, for lower-mass halos, the total stellar mass fraction is dominated by the central galaxy. Hence, we will ignore the contributions from satellite galaxies to the stellar profile. Compared to previous work on the BCM, we therefore modify the stellar density to be an NFW profile with an exponential cut-off:
\begin{equation}
\label{eq:stellar_density}
    \rho_\mathrm{star}(r)\propto \rho_\mathrm{NFW}(r)\exp(-(r/r_\mathrm{cut})^\alpha)\,,
\end{equation}
with the cut-off radius $r_\mathrm{cut}$ and a scaling parameter $\alpha$, controlling the steepness of the transition between NFW and exponential \citep[see e.g.][]{exp_profile_2014}. While we simply use the NFW profile, its outer region does not play a large role due to the exponential cut-off and the parameters chosen. 
\citet{exp_profile_2014} used $1/r$, i.e. a less steep profile than used in this work. We will discuss the impact of these choices on our results in Sec.~\ref{sec:results}.

In the modelling so far, only a single halo was considered for the density profile. However, the density at a given halo position is also influenced by correlated neighbouring halos. This so-called two halo term \citep[see][for a review of the halo model]{asgarii} can be calculated as follows:
\begin{equation}
    \rho^\mathrm{2h}_\mathrm{gas}(r) = \int_0^\infty \frac{k^2\mathrm{d}k}{2\pi^2}j_0(kr) P_{\mathrm{h},\mathrm{gas}}(k)\,,
\end{equation}
with the spherical Bessel function $j_0(x)$ and the \review{gas-halo power spectrum}, $P_{\mathrm{h},\mathrm{gas}}(k)$. Assuming linear halo bias, $b(M)$, the latter is given by
\begin{equation}
\begin{split}
    P_{\mathrm{h,gas}}(k) = & \,b(M)P_\mathrm{lin}(k)\\ & \, \times\int_0^\infty\mathrm{d}M^\prime n(M^\prime)b(M^\prime) u_\mathrm{gas}(k)\,.    
\end{split}
\end{equation}
$u_\mathrm{gas}(k)$ is the Fourier transform of the gas density profile, $n(M)$ the halo mass function and $P_\mathrm{lin}(k)$ the linear matter power spectrum. The total gas profile is then:
\begin{equation}
     \rho_\mathrm{gas}(r) \mapsto  \rho^\mathrm{2h}_\mathrm{gas} + \rho_\mathrm{gas}(r)\,.
\end{equation}
All parameters of the BCM varied in this work are summarised in \Cref{tab:parameters}.

\begin{figure}
\begin{center}
      \begin{tikzpicture}
  \def\r{3}
  \draw (0, 0) node[circle, fill, inner sep=1] (orig) {} -- (-\r/3, \r/2) node[circle, fill, inner sep=1.8, label=above:FRB] (a) {};
    \draw (0, 0) node[circle, fill, inner sep=1] (orig) {} -- (-\r/6, \r/4) node[inner sep=1.8, label=below:$r$] (b) {};
\draw[dashed] (orig) -- (-\r/3, -\r/10) node (phi) {} -- (a);
  \draw (orig) circle (\r);
  \draw[dashed] (orig) ellipse (\r{} and \r/3);
     \draw[->] (orig) -- ++(-\r/5, -\r/3) node[below] (x1) {$x_1$};
  \draw[->] (orig) -- ++(\r, 0) node[right] (x2) {$x_2$};
  \draw[->] (orig) -- ++(0, \r) node[above] (x3) {$x_3$};
    \draw[dashed] (a) -- ++(5.\r, 0) node[right] (observer) {$\mathcal{O}$};
  \pic [canvas is xz plane at y=0,draw=black, text=black, <-, "$\phi$", angle eccentricity=0.55, angle radius=.8cm] {angle = phi--orig--x1};
  \pic [draw=black, text=black, ->, "$\theta$", angle eccentricity=0.6, angle radius=1.35cm] {angle = x3--orig--a};
\end{tikzpicture}
    \caption{Geometry of the model for the DM host contribution as used in \Cref{eq:dm_of_position} expressed in spherical coordinates. The observer $\mathcal{O}$ sits at infinity, so that the integration is along the $x_2$-axis. }
    \label{fig:geometry}
\end{center}
\vspace{.3cm}
\end{figure}
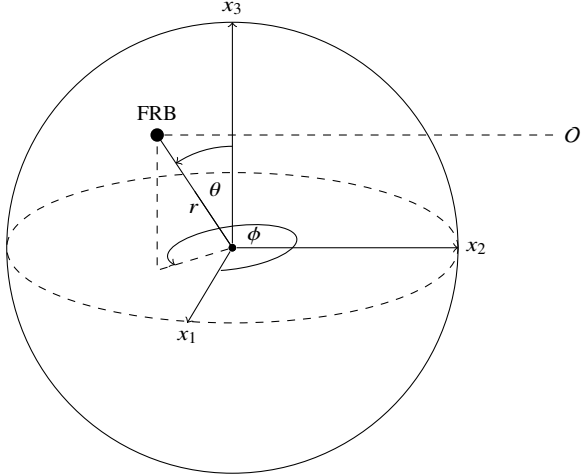

\subsection{Host contribution}
We assume a halo to be described by its mass $M_\mathrm{host}$, redshift $z$ as well as additional parameters such as its concentration and BCM parameters. To ease notation, we collect those in the list $\boldsymbol{\lambda}$. 

The gas density profile of a given halo $\rho_\mathrm{gas}(r;\boldsymbol{\lambda})$ can be related to the comoving electron number density via
\begin{equation}
    \label{eq:electron_number_density}
    n_\mathrm{e}(r;\boldsymbol{\lambda}) = \frac{\rho_\mathrm{gas}(r;\boldsymbol{\lambda})}{\mu_\mathrm{e}m_\mathrm{p}}\,,
\end{equation}
with the proton mass $m_\mathrm{p}$ and the mean molecular weight per electron $\mu_\mathrm{e} = 1.17$. \review{We note that this relation assumes that all baryons are ionised and being present in the interstellar medium. Different measurements indicate $f_\mathrm{IGM}(z) = 90\%\; (80\%)$ at $z \gtrsim 1.5 \; (\lesssim 0.4)$ \citep{meiksin_physics_2009,becker_detection_2011,2012ApJ...759...23S}.}

To compute the host DM, we assume that an FRB is detected inside a host halo at position $\boldsymbol{x} = (x_1,x_2,x_3)^\mathrm{T}$, such that $r= |\boldsymbol{x}|$. Without loss of generality, we align the line-of-sight to the observer, $\mathcal{O}$, located at spatial infinity, with the $x_2$ axis. The geometry of this setting is shown in \Cref{fig:geometry}. In this case, the observed host DM can be computed as
\begin{equation}
\label{eq:dm_of_position}
    \mathrm{DM}(\boldsymbol{x};\boldsymbol{\lambda}) = \int_{x_2}^{x^{\mathrm{max}}_2(x_1,x_3;\boldsymbol{\lambda})} \mathrm{d}x^\prime_2\; n_\mathrm{e}\left(\sqrt{x_1^2 + x^{\prime 2}_2 + x_3^2};\boldsymbol{\lambda}\right)\;,
\end{equation}
where $x^{\mathrm{max}}_2(x_1,x_3;\boldsymbol{\lambda})$ is the maximum value which we consider to be part of the host. 
Throughout this work, we will assume this quantity to be defined via a multiple, $\xi$, of the host halo virial radius:
\begin{equation}
\label{eq:maximum_radius}
    x^{\mathrm{max}}_2(x_1,x_3, ;\boldsymbol{\lambda}) \coloneqq \sqrt{\xi^2 r^2_\mathrm{vir}(\boldsymbol{\lambda}) - x^2_1 - x^2_3}\:.
\end{equation}
If not stated otherwise, we will integrate up to $\xi=3$. \review{This choice is merely numerical and as long $\xi > 1$, its specific choice does not influence the results.}

\subsection{Position within the host is unknown}
The probability, $\mathrm{d}P_\mathrm{host}$  of finding an FRB with host dispersion measure DM in a volume element $\mathrm{d}V$ of the halo is:
\begin{equation}
    \mathrm{d}P_\mathrm{host}(\mathrm{DM};\boldsymbol{\lambda}) = p_\mathrm{FRB}(\boldsymbol{x};\boldsymbol{\lambda})p(\boldsymbol{x}|\mathrm{DM};\boldsymbol{\lambda})\mathrm{d}V\,,
\end{equation}
where $p_\mathrm{FRB}(\boldsymbol{x})$ is the probability to find an FRB at distance $r$ from the halo's centre and $p(\boldsymbol{x}|\mathrm{DM})$ is the probability that a given $\mathrm{DM}$ is produced at position $\boldsymbol{x}$, which itself is given by
\begin{equation}
p(\boldsymbol{x}|\mathrm{DM};\boldsymbol{\lambda}) = \delta_\mathrm{D}(\mathrm{DM}(\boldsymbol{x};\boldsymbol{\lambda}) - \mathrm{DM})\,,
\end{equation}
with the Dirac distribution $\delta_\mathrm{D}$. Thus, in integral form the host probability distribution function (PDF) is
\begin{equation}
\begin{split}
\label{eq:pdf_host}
    p_\mathrm{host}(\mathrm{DM};\boldsymbol{\lambda}) = & \int_0^{\xi r_\mathrm{vir}} r^2\mathrm{d}r\int_0^\uppi\sin(\theta)\mathrm{d}\theta\int_0^{2\uppi}\mathrm{d}\phi  \\
    &\times p_\mathrm{FRB}(r;\boldsymbol{\lambda}) p(r,\theta,\phi|\mathrm{DM};\boldsymbol{\lambda})\,,
\end{split}
\end{equation}
where we assumed that $p_\mathrm{FRB}(\boldsymbol{x};\boldsymbol{\lambda})$ depends on the radius, $r$, only. The distribution of FRBs within their host is still rather uncertain as their origin is still unclear \citep{petroff_fast_2019}, ranging from Magnetars \citep{thornton_population_2013,bochenek_fast_2020} to mergers \citep{Liu_2016}. However, as discussed in \citet{theis_galaxy_2024}, a sensible choice for the placement of FRBs in hosts is to sample them proportional to the stellar mass or star formation rate as they have been identified with Magnetars and globular clusters \citep{kirsten_repeating_2022}. We therefore choose 
\begin{equation}
    p_\mathrm{FRB}(r;\boldsymbol{\lambda}) \propto \rho_\mathrm{star}(r,\boldsymbol{\lambda})\,.
\end{equation}
 In order to compute the host PDF, we note that it is computationally more efficient to sample different sightlines from $p_\mathrm{FRB}(r;\boldsymbol{\lambda})$, compute \Cref{eq:dm_of_position} for all those sightlines and create a kernel density estimate of the resulting histogram. We verified that the sampling approach gives results consistent with the analytic calculation if enough, typically $N\approx 10^4$, sightlines are sampled.

\subsection{Position within the host is known}

Suppose, on top of finding the host, it is possible to locate the FRB inside the host at position $\boldsymbol{x}_\mathrm{FRB}$: 
\begin{equation}
\boldsymbol{x}_\mathrm{FRB} = \begin{pmatrix}
    x_1 \\ x_3
\end{pmatrix} = D_\mathrm{ang}(\boldsymbol{\lambda}) \begin{pmatrix}
    \varphi_1 \\
       \varphi_3
\end{pmatrix}\,,
\end{equation}
here $D_\mathrm{ang}(\boldsymbol{\lambda})$ is the angular diameter distance and $\varphi_{1,3}$ are the angular positions inside the galaxy. We further assume that there is no information about the FRB's line-of-sight position within the host. Thus, one can simply modify $p_\mathrm{FRB}(\boldsymbol{x};\boldsymbol{\lambda})$. In particular:
\begin{equation}
    p_\mathrm{FRB}(\boldsymbol{x};\boldsymbol{\lambda}) \propto\mathcal{P}(\boldsymbol{x}_\mathrm{FRB}, \boldsymbol{{\mathrm{C}}}_\varphi) \rho_\mathrm{star}(r,\boldsymbol{\lambda})\,,
\end{equation}
where $\boldsymbol{{\mathrm{C}}}_\varphi$ encodes the location uncertainties within the host and $\mathcal{P}$ is the corresponding distribution. This scenario requires incredible angular resolution $\ll\mathrm{arcsec}$ and is just provided here for completeness.

\subsection{Uncertainty in the mass estimate}
There are several methods to estimate the mass of a galaxy, including gravitational lensing, dynamics, or using empirically-calibrated mass-to-light ratios \citep[see][for a review]{2014RvMP...86...47C}. All these methods can have significant statistical and systematic errors. Gravitational lensing for example suffers from the mass sheet degeneracy and uncertainty in the lens model. Non-circular motions or ambiguities in the inclination angle influence the mass estimates from dynamic methods. Mass-to-light ratios depend on the initial stellar mass function and stellar evolution models. All these uncertainties are then propagated into an integrated quantity, the mass, and are typically of the order of a few tens of percent. In order to include host halo mass uncertainties in our model, we thus modify Eq.~\ref{eq:pdf_host} and marginalise over host halo mass according to
\begin{equation}
    p_\mathrm{host}(\mathrm{DM};\boldsymbol{\lambda}, M_\mathrm{obs}) = \int_0^\infty \mathrm{d}M \;p_\mathrm{host}(\mathrm{DM};\boldsymbol{\lambda},M) p(M|M_\mathrm{obs})\,,
\end{equation}
where $M_\mathrm{obs}$ denotes the observed mass and $p(M|M_\mathrm{obs})$ is the probability that a halo of that mass has a halo mass $M$. \review{Typically, mass estimates follow a log-normal form distribution, i.e.
\begin{equation}
    p(M|M_\mathrm{obs}) = \frac{1}{M\sigma_M\sqrt{2\uppi}}\exp\left(-\frac{(\ln x - \mu_\mathrm{obs})^2} {2\sigma^2_M}\right)\,,
\end{equation}
such that
\begin{equation}
    M_\mathrm{obs} = \exp\left(\mu_\mathrm{obs} + \frac{\sigma^2_M}{2}\right)\,.
\end{equation}
This is driven by the log-normal scatter in the mass-observable relation \citep[see e.g. in][]{2024PhRvD.110h3509B}. In what follows, we will compare with simulations, where the mass is exactly known. The effect of the uncertainty of the observed mass will therefore not be considered here but needs to be accounted for when confronting the model with real observations.}

\begin{figure}
    \centering
    \includegraphics[width = 0.4\textwidth]{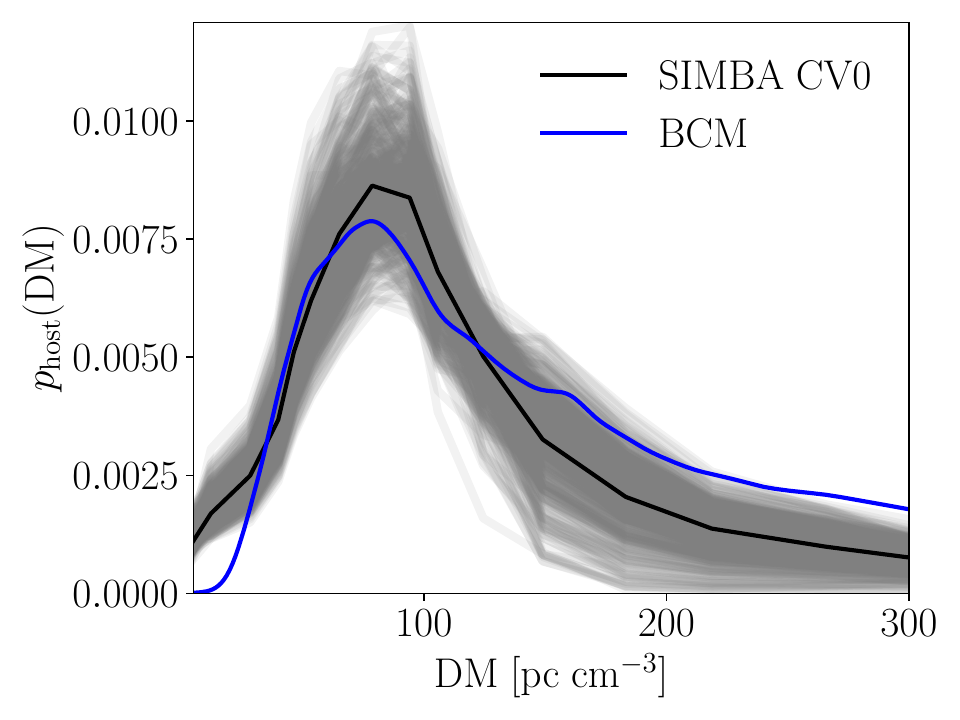}
    \includegraphics[width = 0.4\textwidth]{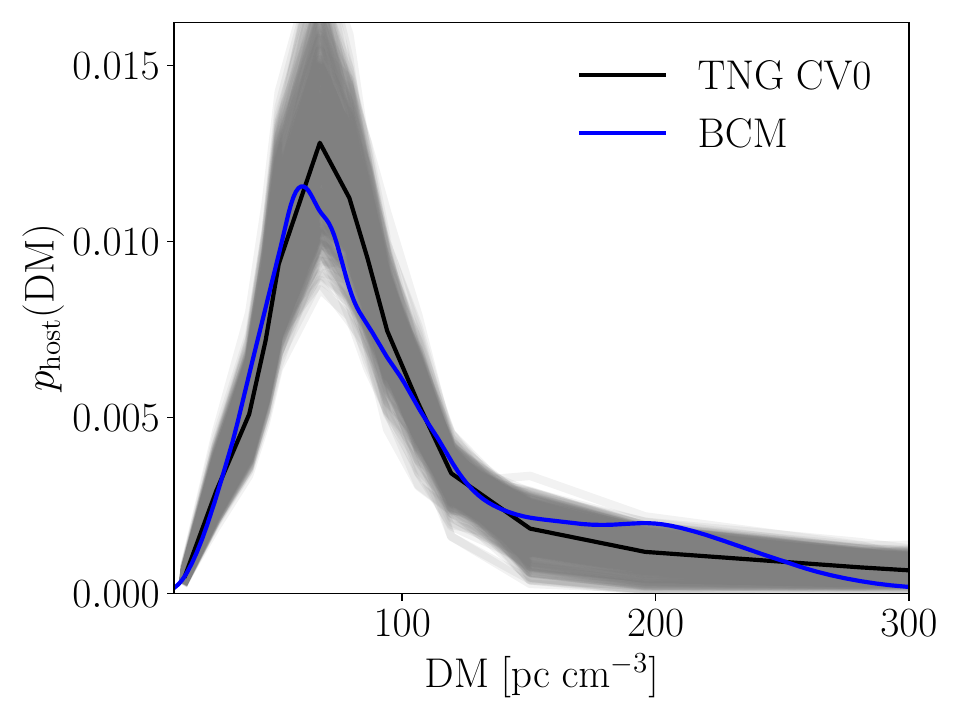}
    \caption{\review{PDF of the DM host contribution in blue with the parameters taken from \Cref{tab:parameters}. 
    The black line corresponds to the PDF measured in SIMBA (top) and IllustrisTNG (bottom) \citep{theis_galaxy_2024} for the cosmic variance realisation CV0. The grey lines are realisations created by drawing Poissonian random numbers from the CV0 halo distribution \citep[see Figure 6 in][]{theis_galaxy_2024}. These are then propagated to the resulting PDF via \Cref{eq:pdfhydro}, creating realisations of the host DM pdf which in turn serves as an error estimate. Here we only show the $68\%$ contours.}} 
    \label{fig:pdf_dm_SIMBA}
\vspace{.3cm}\end{figure}

\begin{figure}
    \centering
    \includegraphics[width = 0.4\textwidth]{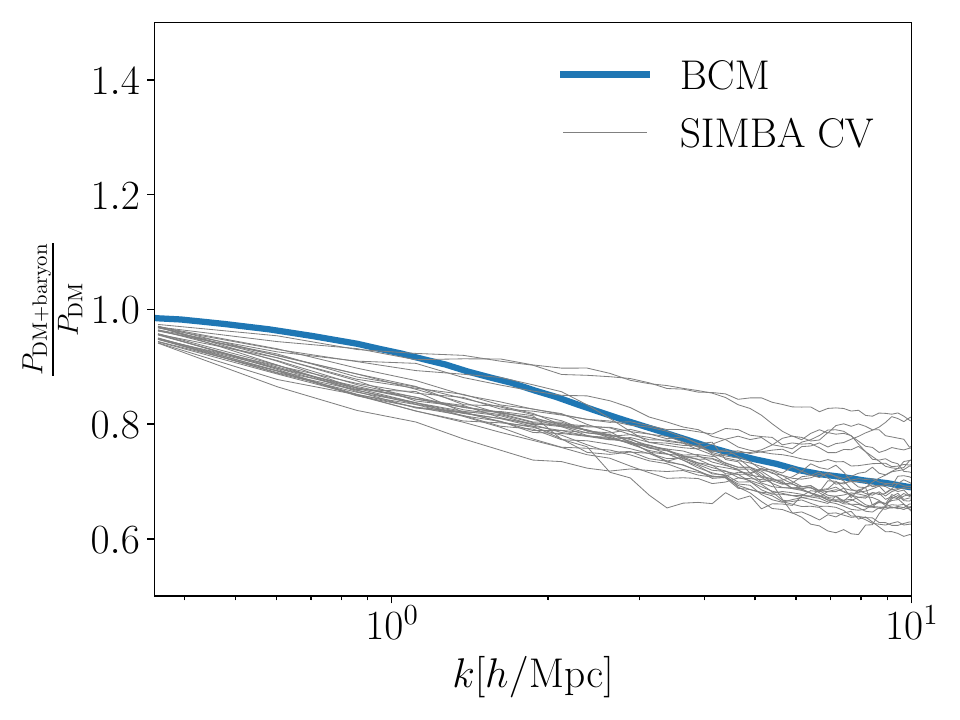}
    \includegraphics[width = 0.4\textwidth]{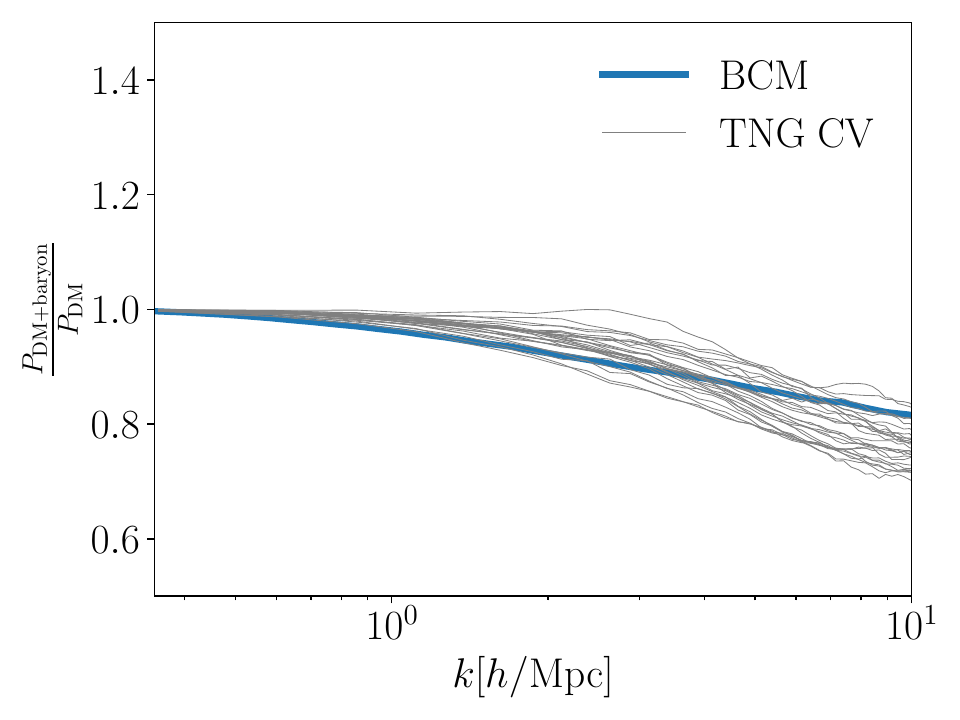}
    \caption{Power spectrum suppression for the parameters from \Cref{tab:parameters} in blue using the emulator from \citet{giri_emulation_2021}. The grey lines correspond to the power spectrum suppression in the different cosmic variance runs of CAMELS. The top plot shows the results for SIMBA, while the lower plot corresponds to IllustrisTNG.}
    \label{fig:sup_SIMBA}
\vspace{.3cm}\end{figure}

\section{Comparison with simulations}
\label{sec:sim}
In order to test and validate our analytic model for the host DM PDF, we first compare it to results from hydrodynamic simulations. In particular, \citet{theis_galaxy_2024} investigated the PDF of the DM host contribution using the Cosmology and Astrophysics with MachinE Learning Simulations \citep[CAMELS,][]{villaescusa-navarro_camels_2021}. CAMELS is a large simulation suite, which currently features 10$\,$680 simulations, with 5$\,$516 of these incorporating hydrodynamic models. These simulations follow the evolution of $256^3$ dark matter particles and $256^3$ gas particles within a comoving volume of $(25 \;h^{-1}\mathrm{Mpc})^3$, beginning at redshift $z = 127$ and providing snapshots from $z = 15$ to $z = 0$. The hydrodynamic simulations in CAMELS are organised into different suites, each representing galaxy formation models based on unique subgrid physics, more specifically IllustrisTNG \citep{weinberger_tng_2017,pillepic_tng_2018}, SIMBA \citep{hopkins_gizmo_2015,dave_2019_SIMBA} and ASTRID \citep{bird_astrid_2022,ni_astrid_2022}. 
 The CAMEL simulations vary six parameters: the total matter density parameter $\Omega_\mathrm{m}$, the root-mean-square amplitude of density fluctuations $\sigma_8$ and four astrophysical parameters—two each for AGN and SN feedback. The three simulation suites differ in baryonic feedback strength due to different implementations of these subgrid processes: SIMBA is expected to exhibit the strongest feedback, followed by IllustrisTNG. Both these suites are anticipated to produce smoother baryon distributions, as compared to ASTRID, which features the weakest baryonic feedback model. However, it is important to note that the true feedback model remains unknown, and different implementations of sub-grid physics can lead to markedly different results that must be distinguished through data. In other words, while the three sets of simulations used in CAMELS cover a range of different feedback models, it is unclear if the true baryonic feedback strength lies within that range.

 For all three simulation suites, CAMELS offers various sets with different properties. Of particular importance for this work is the so-called cosmic variance set (CV), in which simulations differ in their initial conditions but the cosmological and astrophysical parameters are kept at their fiducial values.

In the following, we will focus on the results obtained using the CAMELS CV set in \citet{theis_galaxy_2024} for IllustrisTNG and SIMBA. For more details on how these were constructed from the simulations, we refer the reader to \citet{theis_galaxy_2024}. 

In a first step, we attempt to match the results for the host DM PDF from SIMBA and IllustrisTNG presented in \citet{theis_galaxy_2024}. Since we do not have cosmic variance error bars on the PDF as the results were only obtained for a single run of the cosmic variance sets (CV0), a proper inference cannot be carried 
out. Therefore, we simply measure the distribution function of halos, $p_\mathrm{halo}(M)$, in SIMBA and obtain the PDF of the host DM from the baryonification procedure via
\begin{equation}
\label{eq:pdfhydro}
    p_\mathrm{host}(\mathrm{DM};\boldsymbol{\lambda}) = \int_0^\infty \mathrm{d} M\;  p_\mathrm{host}(\mathrm{DM};\boldsymbol{\lambda},M) p_\mathrm{halo}(M)\,.
\end{equation}
Next, we match the host DM PDF to a reasonable degree by also respecting the prior range of the BCM emulator described in \citet{giri_emulation_2021}. 
In more detail, we assume an unweighted sum of squares of the difference between the measurements from SIMBA/IllustrisTNG and the BCM prediction for the host DM PDF.
\review{Since there are no estimates of the error, we assume that any error in the different runs is given by the number of halos in each simulation as a function of mass. Consequently, we propagate the uncertainty of $p_\mathrm{halo}(M)$ into $p_\mathrm{host}(\mathrm{DM})$ via \Cref{eq:pdfhydro}. For the former, we assume a Poisson scatter around the CV runs, the corresponding histogram can be found in Figure 6 in \citet{theis_galaxy_2024}. We then fit the model, with $\alpha=2$ to the data. For the inference, we assume a Gaussian likelihood and use \texttt{Nautilus} for the parameter estimation \citep{2023MNRAS.525.3181L}.}

\review{The results obtained from this simplistic matching are shown summarised in \Cref{tab:parameters} and the corresponding PDFs are shown in \Cref{fig:pdf_dm_SIMBA}. Since the halo masses in SIMBA/IllustrisTNG are small, $M_\mathrm{c}$ should be in the same range to accommodate variations in $\beta$ with mass. This can be seen by the fact that for $M \ll M_\mathrm{c}$, $\beta \to 0$ irrespective of $\mu$, as can be seen from \Cref{eq:beta}. While $\beta$ will still vary with the halo mass, it is in a regime where these changes are not noticeable in the gas profile, as $\partial_\beta\rho_\mathrm{gas} \propto \beta$. 
Having said that, as long as one is considering a single halo mass, or a relatively narrow mass range, $M_\mathrm{c}$, $\mu$ and $\beta$ are perfectly degenerate, and the only parameter that can be effectively constrained is $\beta$. }

\begin{figure}
    \centering
    \includegraphics[width = 0.45\textwidth]{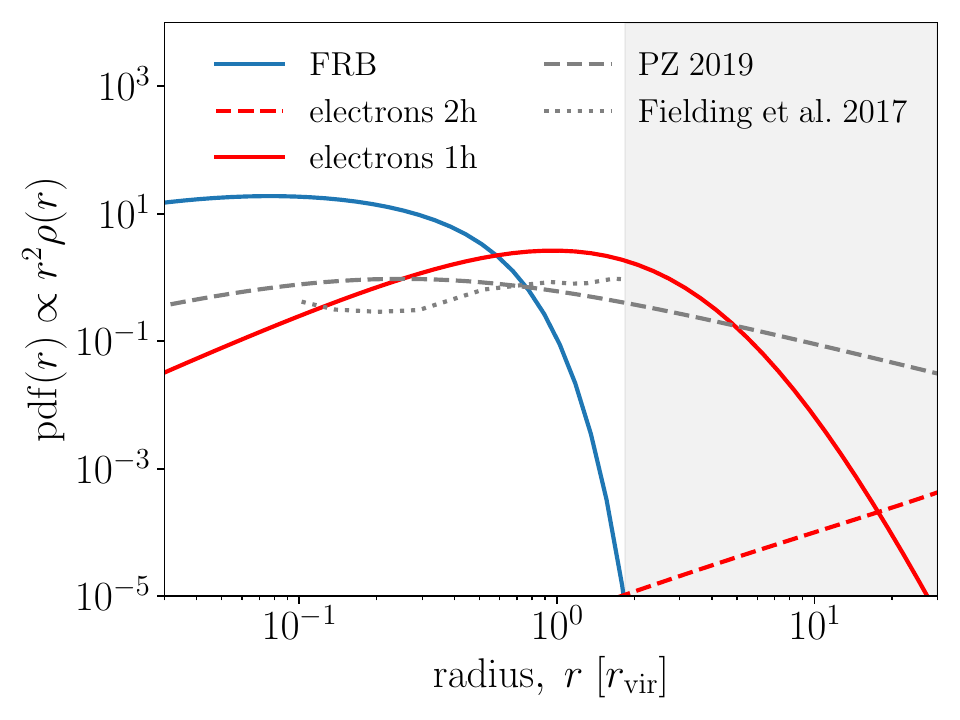}
    \caption{\review{PDF of the electron distribution from the one-halo (1h, red solid) and the two-halo term (2h, red dashed), FRBs (or stars, blue solid), the modified NFW profile from \citet{prochaska_2019} with their fiducial parameters, $\alpha = 2 = y_0$, and the profile from \citet{fielding_2017} in dotted grey. All curves are shown for a halo of viral mass $m_\mathrm{vir} = 10^{12}\;h^{-1}M_\odot$ and at redshift $z=0$ for the fiducial parameters of the BCM. The grey band indicates the range which was not shown in \citet{prochaska_2019}.}}
    \label{fig:pdf_electrons_frb}
\vspace{.3cm}\end{figure}

\begin{figure}
    \centering
    \includegraphics[width = 0.45\textwidth]{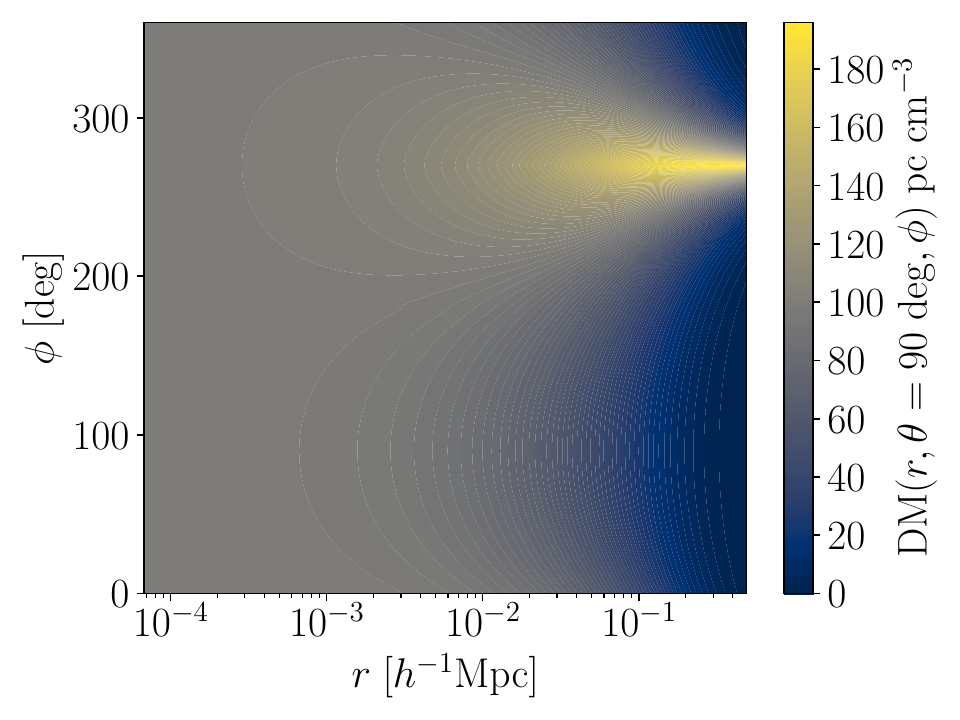}
    \includegraphics[width = 0.45\textwidth]{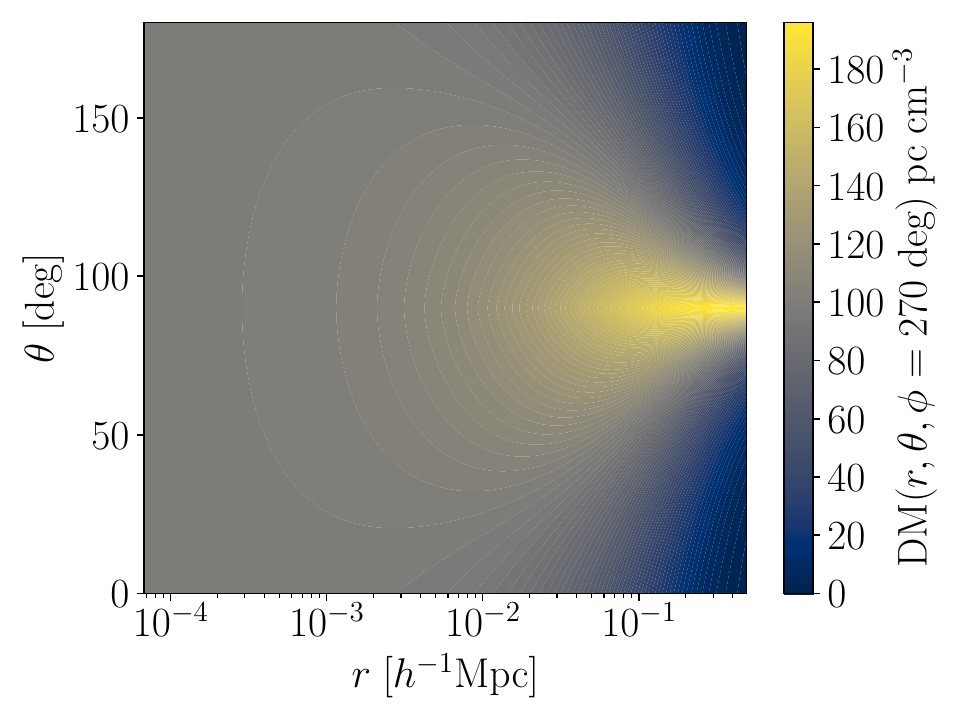}
    \includegraphics[width = 0.45\textwidth]{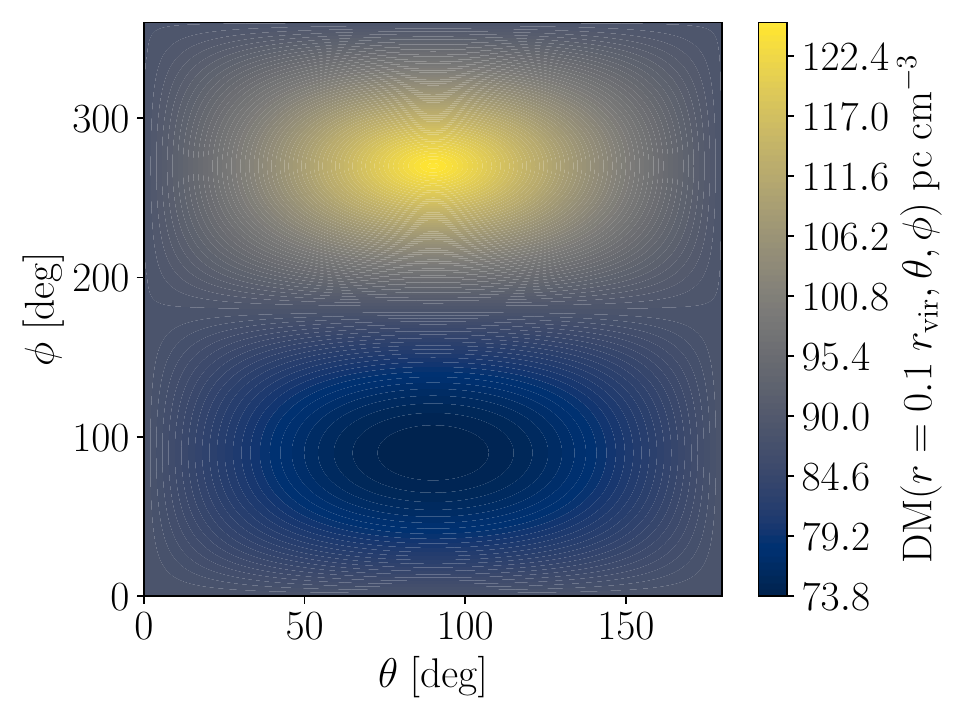}
    \caption{Host DM for different sightlines parametrised in spherical coordinates, standard convention with $\phi$ measured against $x_1$ and $\theta$ with respect to $x_3$. The colour bar shows the DM. Note that the remaining variable is always fixed, as indicated in the colour bar label. \review{The halo has mass is the mean halo mass of TNG $M \approx10^{12}h^{-1}M_\odot$ at redshift zero.}}
    \label{fig:dm_of_position}
\vspace{.3cm}\end{figure}

\review{As can be seen from \Cref{fig:pdf_dm_SIMBA}, the BCM can describe the PDF measured in SIMBA and IllustrisTNG quite well, with the peak and long tail of the distribution clearly reproduced. We can see however, that the tail in the case of SIMBA seems to be overestimated slightly. 
The black line shows the CV0 run of both simulations and the grey shaded area is comprised of Poisson realisations of the underlying halo distribution, $p_\mathrm{halo}$, and propagated into a resulting PDF via \Cref{eq:pdfhydro} using the BCM. This was done because the model is, provided that the FRB sampling distribution and the electron density profile are given, deterministic. Therefore, the only important uncertainty in the simulation is the number distribution of halos in mass. While this shows that the BCM model is flexible enough to fit the PDF of SIMBA and IllustrisTNG, it is important to test if the recovered best-fit parameter values are consistent with expectations: Following our discussion above, it is more instructive to look at $\beta$, i.e. the inner slope of the gas profile, instead of $M_\mathrm{c}$ and $\mu$. For SIMBA, we find $\beta\approx 1.3 $ while IllustrisTNG prefers a smaller $\beta \approx 0.3$ and therefore shallower gas profile in the centre. Inspecting the other parameters in \Cref{tab:parameters}, we see that the ejection radius for SIMBA is larger than the one for IllustrisTNG. This result is indicative of a higher feedback strength in SIMBA, consistent with previous findings \citep[see for example Figure 4 in][]{villaescusa-navarro_camels_2021}.}

As a further consistency test of our model, we use the values of the BCM parameters estimated from the DM host PDF to predict the matter power spectrum suppression using the BCM emulator provided by \citet{giri_emulation_2021}. The results are shown in \Cref{fig:sup_SIMBA}. It should be noted that the value of $\theta_\mathrm{co}$ is not a free parameter of the BCM emulator, and we thus fixed it to 0.1.
This, however, should have minimal influence on the power spectrum directly, as it is only relevant for the smallest scales \citep{schneider_baryonic_2019}.
The resulting suppression is shown in blue. Grey lines indicate the different cosmic variance (CV) runs of SIMBA and IllustrisTNG within CAMELS in the upper and lower Figure, respectively, and therefore give an error estimate on the suppression. It can be seen that the BCM predictions lie within the CV range over all scales estimated in CAMELS. 

These results show that BCM is able to self-consistently fit two different observables within the SIMBA and IllustrisTNG simulations in CAMELS. It should be noted however, that the small box sizes in CAMELS suppress the formation of large halos, and the halo mass range determining the shape of the matter power spectrum in these simulations is thus significantly different from the mass range affecting the true power spectrum: the mean mass of the SIMBA halos used in the comparison above is around $7\times 10^{11}h^{-1}M_\odot$ (roughly 30 percent larger for IllustrisTNG) while e.g. cosmic shear is most sensitive to halos with masses around $10^{13.5}h^{-1}M_\odot$ (for a detailed discussion of the halo mass ranges affecting the matter power spectrum, we refer the reader to Ref.~\citet{2024MNRAS.528.4623V}). Therefore, it is not a priori clear how our results translate to these more realistic cases.

Keeping these caveats in mind, we conclude that the BCM employed here appears self-consistent, i.e. it can reproduce the host DM PDF obtained from two sets of CAMELS simulations while also producing a sensible matter power spectrum suppression. 

\review{Lastly, we note that our formalism is entirely specified by the density profiles of the FRB population, i.e. the stellar density, and the gas density profile. In other words, if uses matches spherically symmetric profiles from simulations to the BCM, the resulting host DM PDFs agree.}

\section{Influence of halo attributes on the DM}
\label{sec:results}
After having established that our model can match the results seen in hydrodynamic simulations for the host DM PDF while at the same time giving a reasonable matter power spectrum suppression, we now use it to better understand what determines the shape of the host DM PDFs.

\review{
In the remainder of the paper, we will now use the BCM parameter values derived from the SIMBA fit and given in \Cref{tab:parameters} as the reference. In addition, we will fix the halo mass to the mean halo mass in TNG,  $M =10^{12}h^{-1}M_\odot$, and the redshift to $z= 0$, unless otherwise stated. Since the halo mass is fixed, the PDFs will look different than the averaged versions shown in \Cref{fig:pdf_dm_SIMBA}.}

\subsection{Profiles and two-halo term}
\review{In \Cref{fig:pdf_electrons_frb}, we show the PDFs, i.e. the respective density profile times $r^2$, of the FRBs (stars in blue) and the electrons (gas in red). The solid red line shows the one-halo contribution, while the dashed line indicates the two-halo term. Note that all profiles have been normalised to the total profile, e.g. $\rho^\mathrm{2h}_\mathrm{gas} + \rho_\mathrm{gas}(r)$. The grey dashed line displays the modified NFW profile studied in \citet{prochaska_2019} which has a very similar shape in the central region but is more extended in the outskirts of the halo. Interestingly, this occurs in a region where the two halo term of our model becomes important, but this was outside the range shown in that paper. We also show the corresponding line from \citet{fielding_2017} in dotted grey which also shows qualitative agreement with the model presented here.} By construction, the FRBs follow the stellar density in a halo and are therefore significantly more concentrated than the electrons, whose distribution extends to much larger radii. We will see later that the difference between the two profiles is the driving factor determining the shape of the PDF of the host contribution. This means that the host halo PDF in principle encodes information about baryonic feedback in this small-scale, low-mass regime.
It can furthermore be seen that on radii $r > r_\mathrm{vir}$ the two-halo term starts to become important as it provides an almost constant density while the one-halo term falls off rapidly. While the integration in \Cref{eq:dm_of_position} is carried out up to a fiducial value of three ($m=3$) virial radii, we show in appendix \ref{app:two_halo} that the influence of the two-halo term on the host DM PDF is negligible. We will therefore ignore the two-halo term for the remainder of this work.

\begin{figure}
    \centering
    \includegraphics[width = 0.45\textwidth]{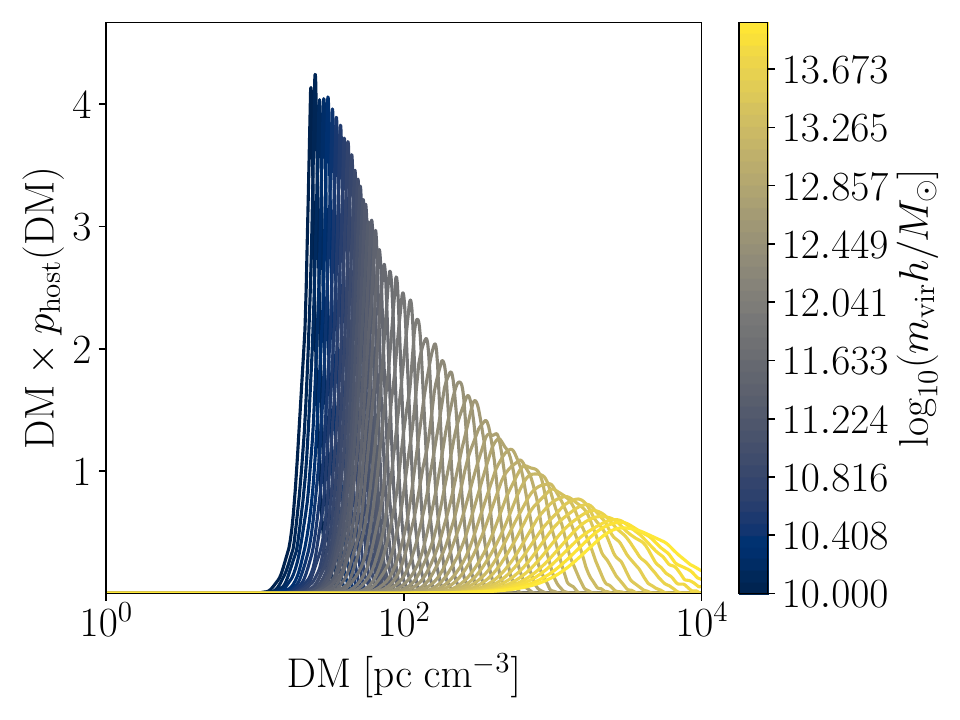}
    \caption{Mass dependence of the host DM PDF, mass ranging from $1\times10^{10}$ to $5\times 10^{14}\;h^{-1}M_\odot$ at redshift $z= 0$.}
    \label{fig:host_mass}
\vspace{.4cm}\end{figure}

\subsection{Dependence on FRB position}
\Cref{fig:dm_of_position} shows the dependence of the host DM on the position of the FRB within the host, parameterised by $(r, \phi, \theta)$ as depicted in \Cref{fig:geometry}.
The first plot shows the $\phi,r$-plane with $\theta = 90\,\mathrm{deg}$. As expected, the DM increases with decreasing radius for $\phi < 180\,\mathrm{deg}$, as the FRB is closer to the centre and the pulse has to pass more electrons along the line-of-sight. For $\phi > 180\,\mathrm{deg}$ the trend is, however, reversed since the pulse will see additional electrons from the trailing edge of the halo on top of those located at the leading edge. The largest DM can be seen at $\phi = 270\,\mathrm{deg}$, i.e. when the FRB is directly aligned with the symmetry axis.
In the second panel, $r$ and $\theta$ are varied with $\phi = 270\,\mathrm{deg}$. We note the symmetry at $\theta =90\,\mathrm{deg}$ as expected. Furthermore, we can again see that the DM increases with increasing radius due to the FRB being located in the trailing half of the halo. This is strictly only true if the FRB is located in the plane of $\theta = 90\,\mathrm{deg}$. For different values, this behaviour can be reversed.
Additionally, the DM also decreases with $\theta$ decreasing or increasing from $\theta= 90\,\mathrm{deg}$. Finally, in the last panel the radius is fixed to $r = 0.1r_\mathrm{vir}$ and the $\theta,\phi$-plane is shown, displaying once again the strong dependence on the polar angle.

\begin{figure}
    \centering
    \includegraphics[width = 0.45\textwidth]{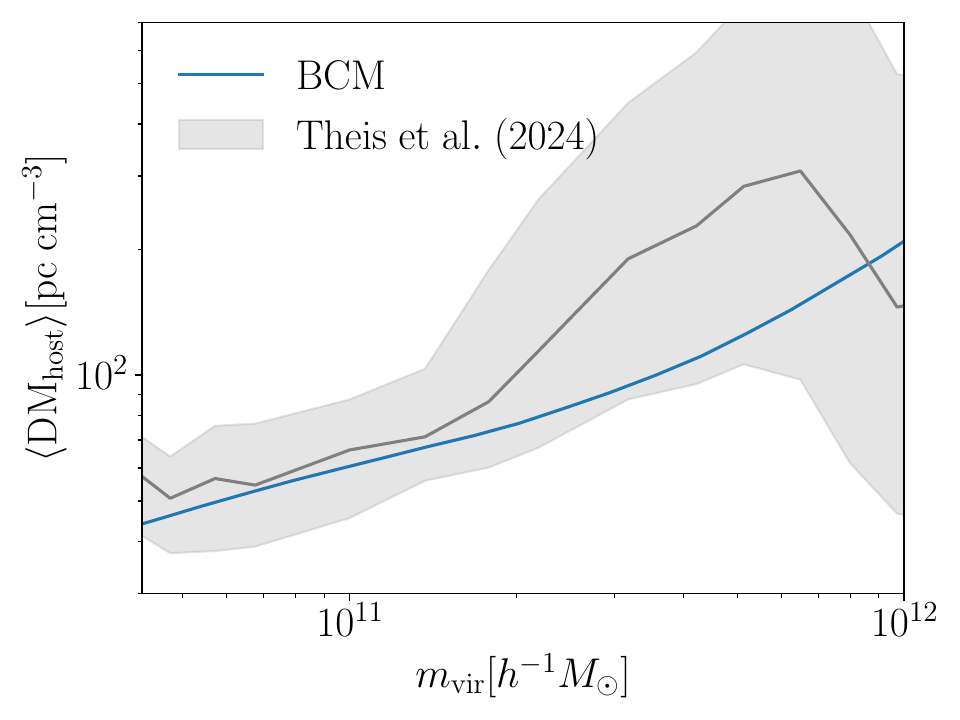}
    \caption{Mean host DM as a function of halo virial mass $m_\mathrm{vir}$ from the BCM in blue. The grey shaded area indicates the scatter in the host DM obtained from \citet{theis_galaxy_2024}.}
    \label{fig:host_mass_theis}
\vspace{.3cm}\end{figure}

\subsection{Mass dependence}

Our analytic model also allows us to investigate the dependence of the host DM PDF on halo mass. In \Cref{fig:host_mass} we show the corresponding distributions for halo masses between $9\times 10^{10}h^{-1}M_\odot$ and $5\times 10^{12}h^{-1}M_\odot$ as colour-coded. For all cases, the redshift of the halo is set to zero. From the plot, it is clear that the average host DM increases with halo mass. At the same time, the distribution also becomes broader and tends to become more log-normal. This leads to very long tails in DM, especially at high masses. Recently, this has been observed in zoom-in simulations \citep{2024arXiv240603523O} where substantial DMs were found in extreme edge-on cases (as we saw in \Cref{fig:dm_of_position} as well). These results might indicate that the mean host contribution and its scatter are larger than previously assumed  \citep[e.g. in][]{hagstotz_new_2022,reischke_probing_2021}. If confirmed, this could have consequences on the inference of the cosmological background model with FRBs, since the host contribution might be more dominant and introduce a larger uncertainty due to its scatter. In case of a measurement of DM correlations, the increased scatter could enhance the observed noise levels in the correlation function measurement, making DM correlations harder to detect. Thus, more FRBs than previously assumed would be required to make such a detection.

\review{
To compare the mass scaling directly to CAMELS, we again use the results from \citet{theis_galaxy_2024}: Specifically, in \Cref{fig:host_mass_theis} we compare the mean host DM as a function of host halo mass as measured from SIMBA to the predictions using the best-fit BCM parameters derived in the previous section (see \Cref{tab:parameters}). The grey-shaded region corresponds to the results from CAMELS, with width determined by the 68 percent contour of the host DM PDF. Comparing our predictions to the simulation results, we can see that the general trend is reproduced quite well.
Looking at \Cref{fig:host_mass_theis} the main difference between the analytical model and the numerical simulations is that the DM is slightly lower on average in the analytical model at low masses. Furthermore, the simulations show some non-monotonicity in the mean host DM, i.e around $8\times 10^{12}h^{-1}M_\odot$, the mean host DM seems to decrease. As discussed in \citet{theis_galaxy_2024}, this could be due to the limited number of halos at high masses since the associated Poisson noise is not accounted for in this Figure. An alternative possibility are the feedback mechanisms in the simulations themselves where some processes become important only above a specific mass threshold. This can lead to more baryons thrown out of halos and therefore to a lower absolute baryon content even if the halo mass increases. While the BCM model has a non-monotonic functional form for the fractional abundances, see \Cref{eq:star fraction}, the absolute gas content in each halo is monotonic with the halo mass. Hence, this non-monotonic feature will not be reproduced by the BCM in its current version}.

\begin{figure}
    \centering
    \includegraphics[width = 0.45\textwidth]{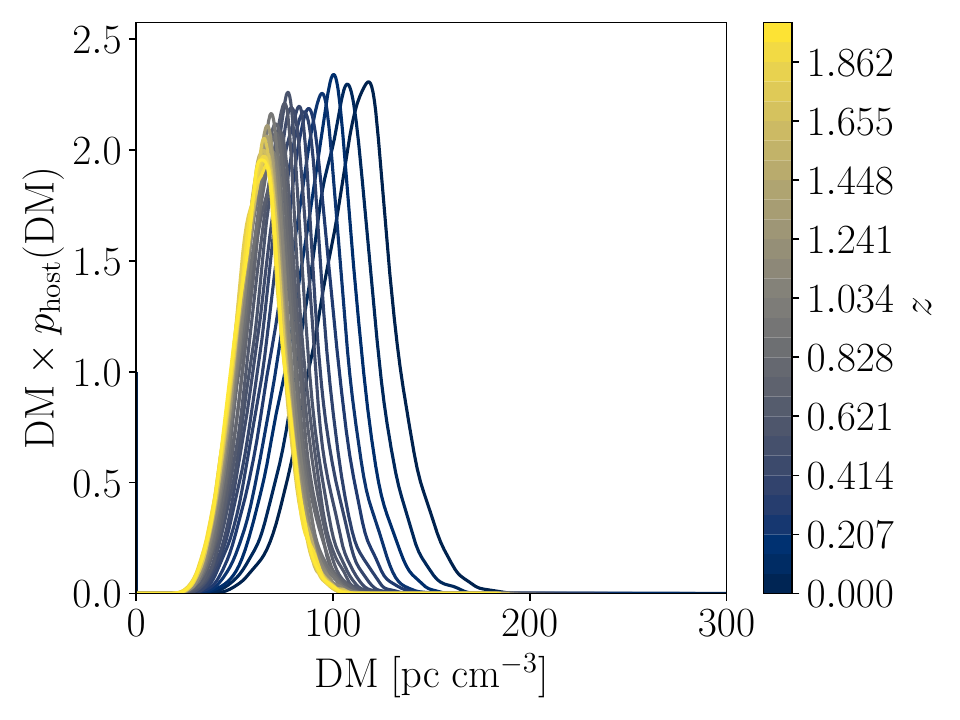}
    \caption{Redshift dependence if the host DM PDF. The mass of the halo is $M= 10^{12}h^{-1}M_\odot$.}
    \label{fig:host_redshift}
\vspace{.3cm}\end{figure}

\subsection{Redshift dependence}
 We investigate the redshift dependence of the host DM PDF in \Cref{fig:host_redshift}. It can be seen that the mean host contribution decreases slightly with increasing redshift, consistent with numerical simulations \citep{theis_galaxy_2024}. Note that this does not include the $(1+z)^{-1}$ factor from the rest frame transformation (see Equation \ref{eq:dm_los}). This means that there is a residual redshift evolution in the host contribution.
 Furthermore, the distribution becomes increasingly compact with increasing redshift. This can be explained by the observation that baryons are less diffuse at higher redshift, due to the reduced time available for redistribution by feedback processes.

The increase of the host DM with decreasing redshift in the simulation is also driven by structure growth. While the star formation rate peaks around a redshift of 2, hierarchical structure growth produces generally more free electrons in halos. 
We want to stress that the intrinsic redshift evolution of the host DM PDF seen in \Cref{fig:host_redshift} does neither mimic the one of the LSS term,  $\mathrm{DM}_\mathrm{LSS}(z)$, nor the $(1+z)^{-1}$ scaling of $\mathrm{DM}_\mathrm{host}$ in \Cref{eq:parts}. Consequently, observing many FRBs with host identification at different redshifts could potentially allow us to tell these different contributions apart, as for example at low redshifts the DM of FRBs will be dominated by the host contribution, while at high redshifts the LSS contribution will take over as it grows with redshift. 

\subsection{BCM parameter dependence}
\begin{figure}
    \centering
    \includegraphics[width=.41\textwidth]{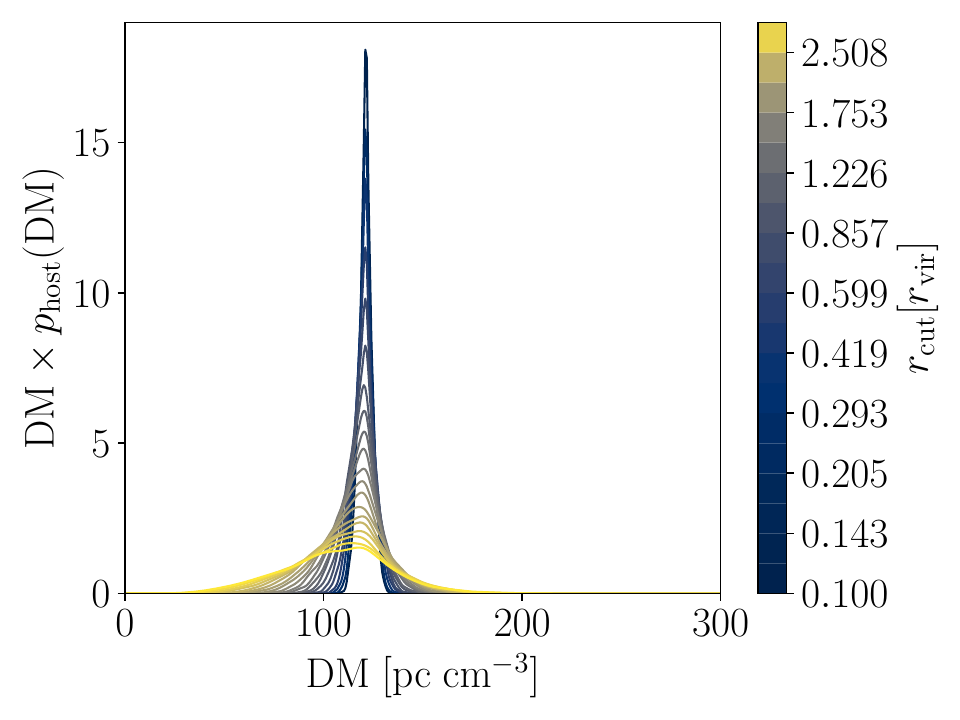}
    \includegraphics[width=.41\textwidth]{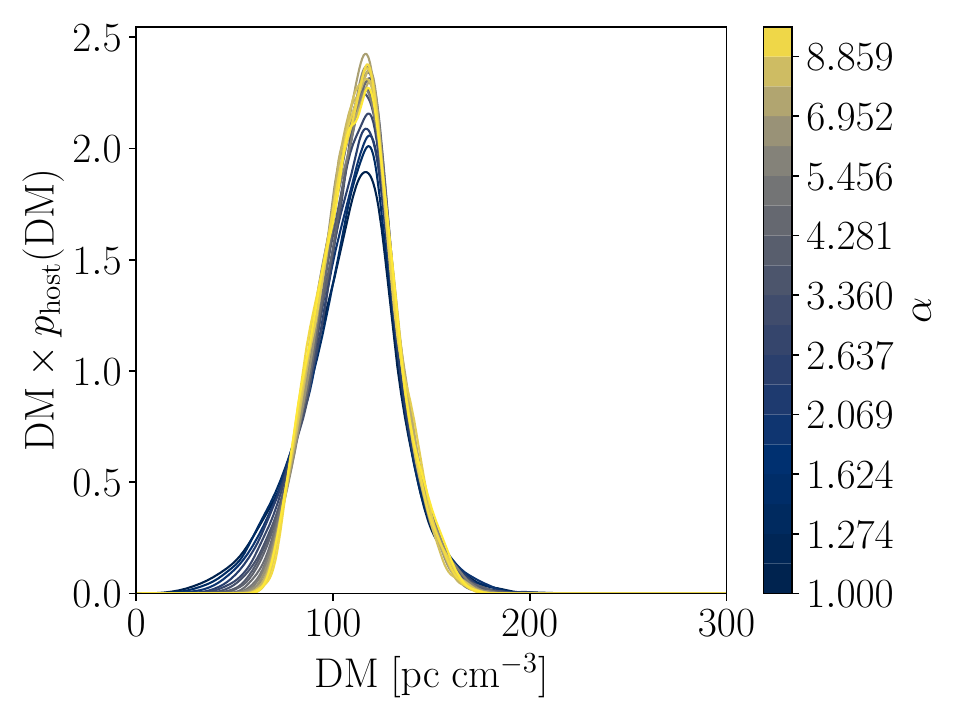}
    \caption{Dependence of the host DM PDF on the BCM parameters (see \Cref{tab:parameters}) of the stellar distribution. For all plots, we fix the halo mass to $M = 10^{12}h^{-1}M_\odot$ and the redshift to $z= 0$. Parameters which are not varied are always set to their respective fiducial values. Note the different DM range in the upper plot.}
    \label{fig:parameter_baryonification_star}
    \vspace{.3cm}
\end{figure}

\begin{figure*}
    \centering
    \includegraphics[width=.33\textwidth]{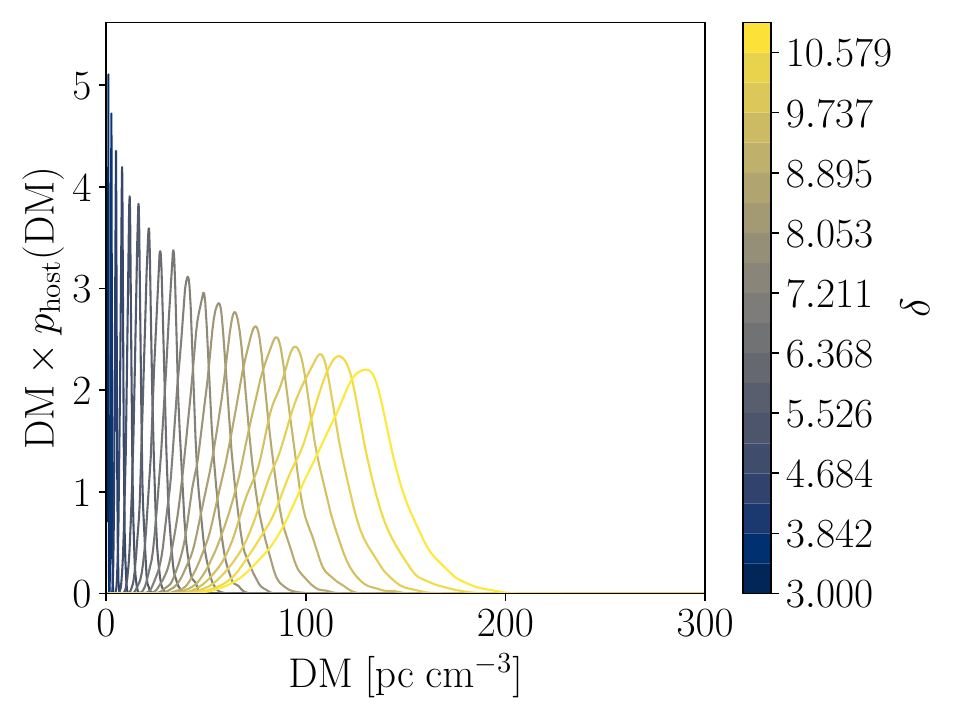}
    \includegraphics[width=.33\textwidth]{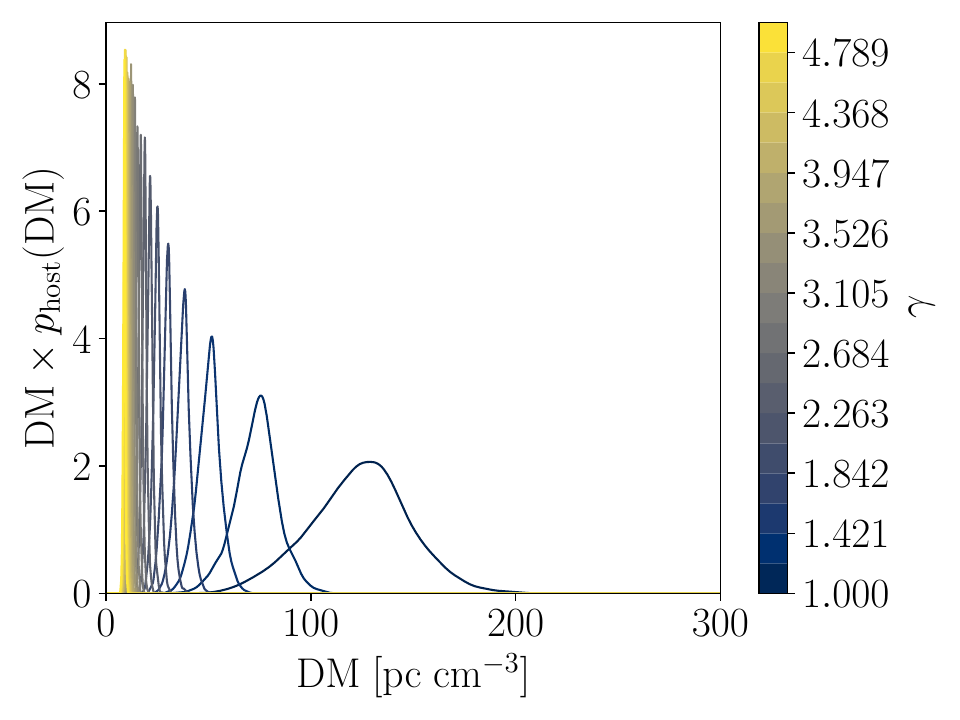}
    \includegraphics[width=.33\textwidth]{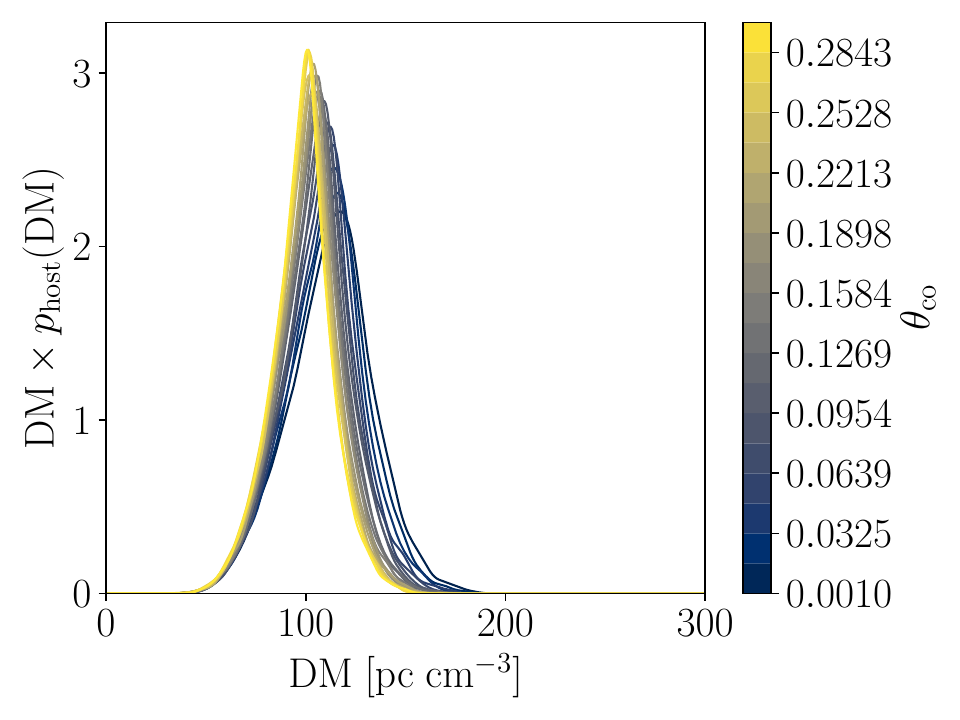}
    \includegraphics[width=.33\textwidth]{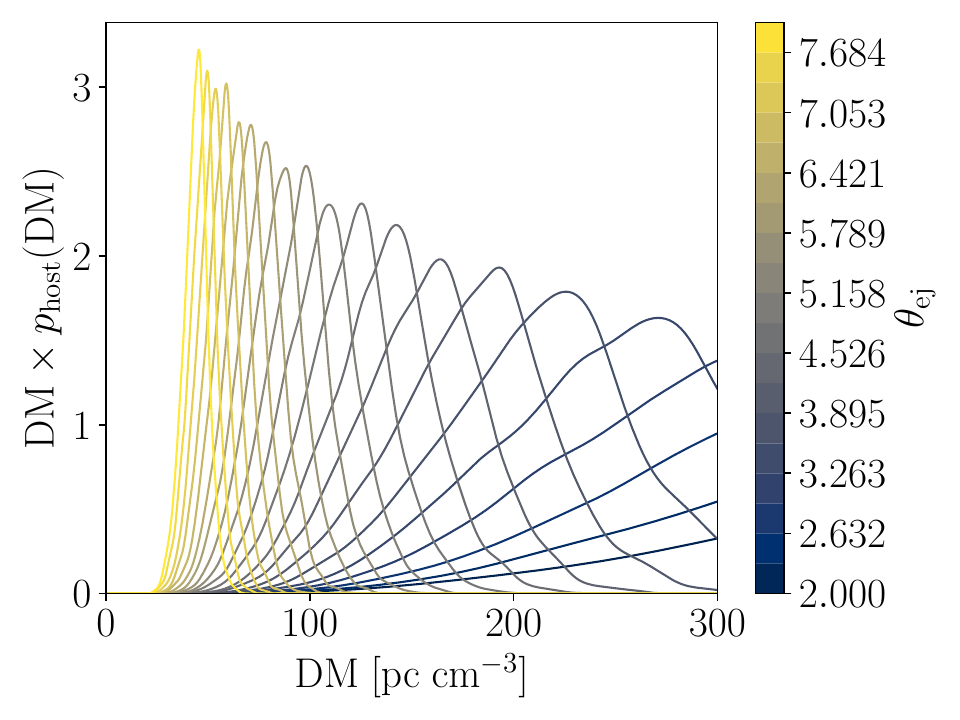}
    \includegraphics[width=.33\textwidth]{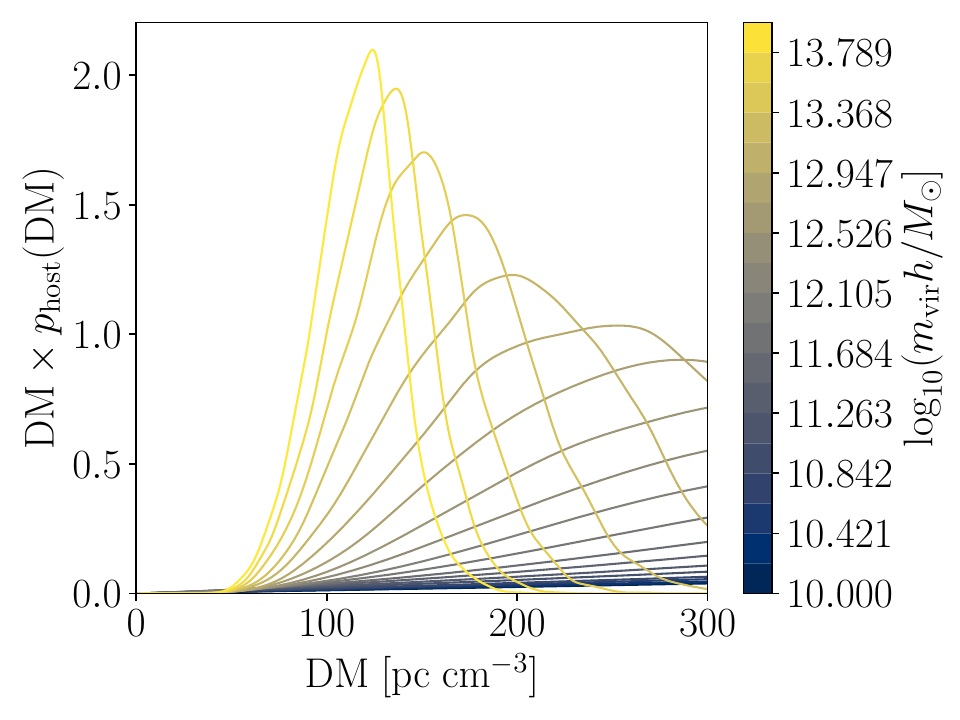}
    \includegraphics[width=.33\textwidth]{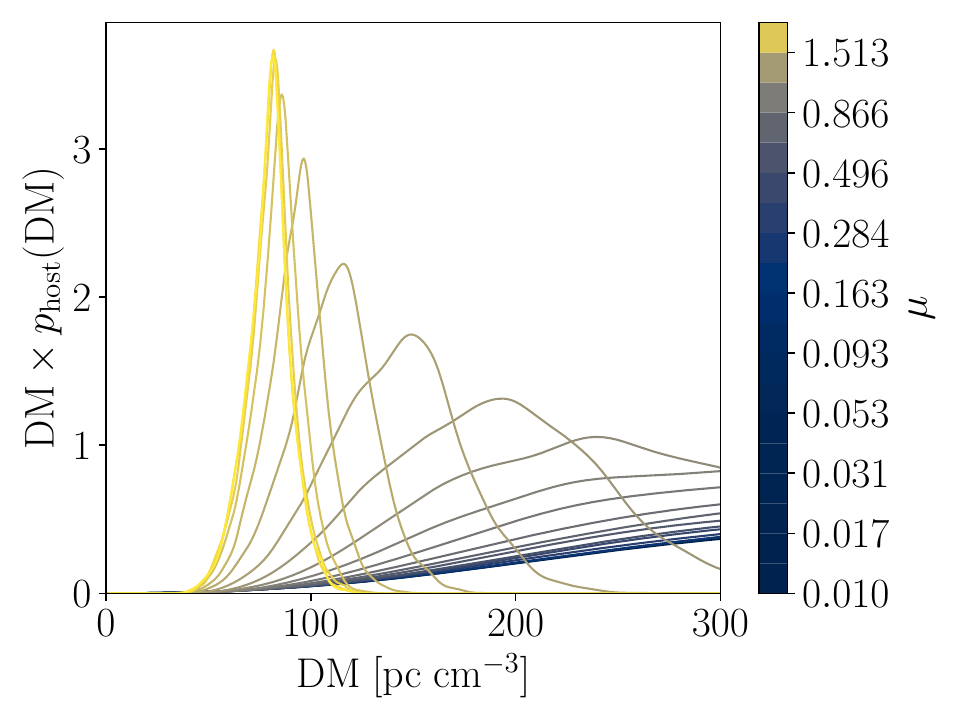}
    \caption{Dependence of the host DM PDF on the BCM parameters (see \Cref{tab:parameters}) of the gas distribution. For all plots, we fix the halo mass to $M = 10^{12}h^{-1}M_\odot$ and the redshift to $z= 0$. Parameters which are not varied are always set to their respective fiducial values.}
    \label{fig:parameter_baryonification}
    \vspace{.3cm}
\end{figure*}

Finally, we use our model to investigate the dependence of the host DM PDF on BCM parameters. The results for all eight BCM parameters described in \Cref{tab:parameters} are shown in \Cref{fig:parameter_baryonification_star} and \Cref{fig:parameter_baryonification}. We vary each parameter separately and fix the remaining ones to their fiducial values. In all cases, the halo mass and the redshift are kept fixed at $M= 10^{12}h^{-1}M_\odot$ and $z =0$.

First, we investigate the dependence on the parameters mainly controlling the stellar distribution. \Cref{fig:parameter_baryonification_star} shows the dependence of the host DM PDF on the exponential cut-off of the stellar density (see Equation \ref{eq:stellar_density}), i.e. the PDF of the FRBs, with $r_\mathrm{cut}$ controlling the position of the cut-off and $\alpha$ the transition from the inner profile to the exponential cut-off. The results for these two parameters are shown on the upper and the bottom panel, respectively.
 Curves with small $r_\mathrm{cut}$ correspond to a profile with a steep exponential cut-off, while large values of $r_\mathrm{cut}$ would result in a pure NFW profile for the sampling of FRBs. In the latter case, the distribution of FRBs is similar to the gas distribution within the halo. Comparing to our results, this suggests an FRB distribution significantly steeper than both the gas and the NFW profiles. This is due to the fact that, as $r_\mathrm{cut}$ decreases, FRBs will originate from a more localised region of the halo's core, and they will thus have higher DMs on average, which exhibit a lower spread in their values. 
 In other words, there will be only very few FRBs with small DM. This can therefore not reproduce the long tail we see in the simulations, which tells us that the stellar profile cannot be too concentrated in the halo's centre.
The dependence on $\alpha$ is less strong. Increasing $\alpha$ leads to slightly higher mean DMs while somewhat reducing the overall width since the FRBs are sampled from a smaller range of radii as discussed above.

 Let us now turn to the parameters determining the shape of the gas profile: In the first plot in the top row of \Cref{fig:parameter_baryonification} we illustrate the dependence on $\delta$ on the left-hand-side. As can be seen, we find the DM PDF to strongly depend on $\delta$, with increasing values producing a broader distribution. This is due to the fact that increasing $\delta$ makes the gas profile more compact, as it increases the steepness of the outer parts and leads to higher central densities due to mass conservation. The first effect pushes the mean to larger DM values, while the latter produces the tail of the host DM PDF. This illustrates a crucial aspect of our model: large DM values are mainly produced by the central part of the profile, since this is also very likely the origin of FRBs due to the stellar density profile. Intuitively, one might think that, even if the electrons are pushed out of the halo, they will still be integrated along the line-of-sight. However, a good fraction of the sampled FRBs will never encounter those outer regions and therefore miss the electrons along the line-of-sight because they have been moved from the inner to the outer part of the halo. Therefore, a high central density is more important for a large average DM than a shallower slope of the outskirts of the gas density. In other words, the volume fraction of the electron density which is passed by an FRB is much lower in the outer parts of the halo than in the inner parts due to the very compact stellar density profiles.
 
 The dependence on $\gamma$, first row in the middle, is opposite to $\delta$, with decreasing $\gamma$ leading to longer tails. $\gamma$ modifies the density profile in the transitional region, mediating between the two power-laws in \Cref{eq:rhogas}. In particular, for $r \ll \theta_\mathrm{ej}r_\mathrm{vir}$, the gas density profile is characterized by the inner slope, $\beta$, while it follows the outer slope $\delta-\beta$ for $r \gg \theta_\mathrm{ej}r_\mathrm{vir}$. Increasing $\gamma$ increases the density of the region between these two regimes, consequently decreasing the density in both the inner and outer region.

\review{The last plot in the first row shows the impact of $\theta_\mathrm{co}$. As can be seen, smaller values of $\theta_\mathrm{co}$ lead to a slightly wider DM distributions with longer tails. This can be understood as follows: If $\theta_\mathrm{co} \to 0$, the inner profile will be characterized by the inner slope $\beta$. In contrast, if $\theta_\mathrm{co} \to \infty$, the inner region of the density profile will flatten. As a flat profile will have a much lower density in the centre, larger values of $\theta_\mathrm{co}$ lead to smaller DM values on average. Furthermore, due to this flatness, there will be minimal variability in the DM regardless of the FRB position, making the PDF very narrow allowing only a small range of possible DM values. Small values of $\theta_\mathrm{co}$ in contrast, lead to high electron densities in the centre, again producing the characteristic long tail in the host DM PDF. This discussion, however, hinges on the value of $\beta$ and hence on the ratio of $m_\mathrm{vir}/M_\mathrm{c}$. For the values discussed here $m_\mathrm{vir}/M_\mathrm{c}< 1$ and in particular $\beta = 0.27$, hence the profile is already fairly flat in the centre, thus changing $\theta_\mathrm{co}$ does not affect the overall gas profile that much. This would be different if we consider a higher halo mass or a different value for $M_\mathrm{c}$.}

For $\theta_\mathrm{ej}$, the situation is similar, but this time the argument applies to the outer region of the density profile. When $\theta_\mathrm{ej} \to \infty$, gas is pushed further out from the halos' centre, while the opposite is true for $\theta_\mathrm{ej} \to 0$. Consequently, due to the higher densities in the central region of the halo at low $\theta_\mathrm{ej}$, we find on average larger DMs and a longer tail in the host DM PDF, as shown in the first plot of the lower row in \Cref{fig:parameter_baryonification}. \review{Similar to the discussion regarding $\theta_\mathrm{co}$ above, the exact dependence of $\theta_\mathrm{ej}$ strongly depends on the values of the remaining parameters.}

The last two plots in \Cref{fig:parameter_baryonification} show the dependence on the two BCM parameters $M_c$ and $\mu$, which determine the inner slope of the gas profile and control its mass dependence. Consequently, if evaluated for an individual halo, one cannot measure both parameters simultaneously due to their degeneracy. In this discussion, we will therefore describe how $\mu$ and $M_\mathrm{c}$ change $\beta$ and in turn the gas profile. 
As already seen, a steeper inner slope corresponds to larger densities in the halo centre, i.e. we expect large $\beta$ to lead to wider distributions of the host DM as well as larger average values.
Since for $M_\mathrm{c} \gg M$, $\beta \to 0$ we expect a very low central electron density in this case. From the central plot in the second row, this effect can be seen clearly: if $M_\mathrm{c}$ is increased significantly above the mass of the halo, small DMs are the result. Conversely, if $M_\mathrm{c} < M$, the central profile becomes steeper, producing larger DMs. 
\review{The dependence on $\mu$, at a fixed halo mass, depends on the ratio of $m_\mathrm{vir}/M_\mathrm{c}$. If $m_\mathrm{vir}/M_\mathrm{c} < 1$, increasing $\mu$ will decrease $\beta$ and vice versa for $m_\mathrm{vir}/M_\mathrm{c}>1$. As an example, let us assume $m_\mathrm{vir}/M_\mathrm{c} <1$ as this is the situation we are presented with here, in this case 
increasing $\mu$, will increase $\beta$. With the arguments presented before, larger DMs are expected with increasing $\mu$ which is exactly what can be seen in the figure.}

\subsection{What does the host contribution measure?}
Due to the large number of parameters discussed, we would like to close this section by summarising the most important takeaways from the dependence of the host DM PDF on the BCM. Steeper profiles in the centre of halos produce larger DM; this is true for small $M_\mathrm{c}$, $\theta_\mathrm{ej}$, $\theta_\mathrm{co}$, and $\gamma$, or large $\mu$ and $\delta$. In general, the long tail of the host DM PDF is produced by a very localised stellar population and a cusp-heavy gas distribution, with the outer parts of the halo only having a smaller influence on the host DM PDF. 
 These two key features, i.e. non-zero average DM and log-tailed distributions, are needed to match results from hydrodynamic simulations \citep[see e.g. ][]{theis_galaxy_2024}.

Lastly, it is instructive to discuss what the host contribution actually measures. From \Cref{eq:pdf_host}, it is clear that the host contribution depends on the product of the gas and the stellar profile. Since the standard BCM parameters are mostly describing the latter, we will discuss the influence of the stellar profile on the inferred gas distribution in this section. The fiducial assumption of the stellar distribution is shown in \Cref{eq:stellar_density}, i.e. an NFW profile which is exponentially cut off. Depending on the chosen cut-off radius, only the inner part of the profile matters, i.e. $\rho_\mathrm{star} \propto r^{-2}$. 

The results shown in \Cref{fig:parameter_baryonification_star} suggest that a decrease in the cut-off radius will generally produce a larger DM on average. At the same time, however, the scatter in the DM reduces as the FRBs are sampled from a smaller spatial region. A larger cut-off radius, on the other hand, always moves the peak of the distribution closer to $\mathrm{DM}=0$. This suggests that there exists a sweet spot between the two regimes if one aims to reproduce hydrodynamic simulations. We investigate this further by assuming a different stellar profile and only fitting the BCM parameters of the hot gas to the simulations. We distinguish three profiles: $(a)$ the fiducial stellar profile, $(b)$ a more compact profile  $r_\mathrm{cut} = 0.1 r_\mathrm{vir}$ and $(c)$ a less compact profile with $r_\mathrm{cut} = 0.5 r_\mathrm{vir}$. 

\begin{figure}
    \centering
    \includegraphics[width = 0.45\textwidth]{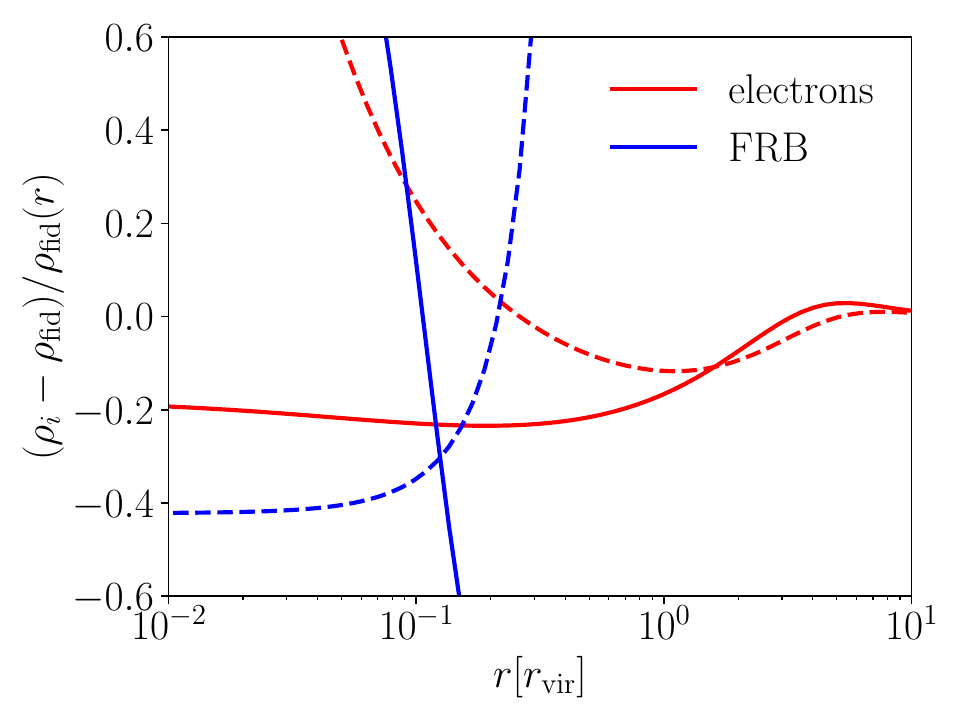}
    \caption{Ratio of the densities of case $(b)$, i.e. a more compact stellar profile, in solid and case $(c)$, i.e. a less compact stellar profile, in dashed relative to the fiducial case $(a)$. The blue lines show the ratio of the stellar profile. Red lines show the ratio of the gas profiles which best match the hydrodynamic simulation (SIMBA in this case).}
    \label{fig:profiles_ratio}
    \vspace{.3cm}
\end{figure}

In \Cref{fig:profiles_ratio} we show the ratio of the stellar and gas profiles relative to the fiducial case ($i$). If the FRBs originate in a smaller region of the halo (solid lines), the BCM reacts to this by pushing more gas out of the central region of the halo into the outskirts to still match the simulation. If the gas profile would not be changed, the average DM would be larger and the long tail of the distribution would vanish (as shown in the left plot of \Cref{fig:parameter_baryonification_star}). This can be understood by considering a limiting case with an infinitely compact FRB distribution. In this case, the PDF would correspond to a Delta-distribution. The shallower gas profile counteracts this effect by reducing the central density while at the same time moving gas outside of the halo so that it can still be traversed by some FRBs to obtain a large DM. 
For a less compact stellar profile, the opposite is true. A larger gas density is required to produce large enough DMs (this is shown in dashed in \Cref{fig:profiles_ratio}).

In conclusion: the long-tailed, quasi-log-normal distributions with non-zero mode DM observed in simulations arise due to the interplay of the FRB and the gas distributions and can generally be reproduced in two different ways: (i) Either with a steep and compact FRB distribution and correspondingly shallower gas density profile, or (ii) with a shallower distribution of FRBs and a more compact gas density. In the parlance of baryonic effects, case (i) would correspond to comparatively stronger feedback and case (ii) to weaker feedback. However, using the DM PDF by itself, we cannot distinguish between these two cases without strong priors on the FRB distribution. It is interesting to point out though that this degeneracy is not perfect, i.e. strong feedback cannot be compensated by increasingly compact stellar distributions, and the same applies in reverse for weak feedback.

\vspace{1cm} 
\subsection{Limitations of the current model}
While the model presented here can self-consistently reproduce different observables in hydrodynamic simulations, it is subject to a number of limitations, which we would like to address in the following. 

First, in our work we have assumed a simplistic model for the stellar (and thus FRB) distribution in halos, and it is unclear if this model provides an accurate description of observations. While other applications of the BCM, e.g. the total matter power spectrum for cosmic shear or the gas density and pressure for the Sunyaev-Zel'dovich effects are not strongly affected by this choice, our results are very sensitive to the stellar profile, as shown in Sec.~\ref{sec:results}. The simplicity of the stellar profile might become even more important at larger redshifts where star formation peaks, and thus presents a significant caveat in our modelling. 

This leads to the next possible limitation: fitting observables in different mass ranges. While we have shown that we can reproduce both the host DM PDF, and the matter power spectrum suppression measured in the CAMELS suite of hydrodynamic simulations, it is not obvious how these results transfer to an extended mass range. For example, for cosmic shear, large halo masses (around $10^{13.5} h^{-1}M_\odot$) determine the shape of the observed matter power spectrum \citep[see e.g.][]{arico_des_2023}, while the host DM PDF will be dominated by low-mass halos, i.e. $M < 10^{12} h^{-1}M_\odot$. This raises the question if a joint fit requires a more complex model with a more complicated mass dependence, as advocated for example in \citet{anbajagane_map_2024,arico_baryonification_2024}. 
A parameter for which a more complex parametrization might be of particular importance is $M_\mathrm{c}$, given the low values required by our analysis as compared to previous BCM analyses \cite[see e.g.][]{arico_des_2023}. We leave an investigation of these questions to future work.

Furthermore, low-mass halos are expected to contain a non-negligible fraction of cold gas \citet[see e.g.][]{tumlinson_cos_2013,decataldo_origin_2024}. Although we expect this cold gas component to not be fully ionized, it could have an effect on the observed host halo DMs. It would thus be interesting to extend the analysis presented here to also include a cold gas component in addition to the hot gas distribution considered.

Lastly, it is also worth mentioning that there are limits to the accuracy of the hydrodynamic  simulations we compared to; in particular, due to the small box sizes of CAMELS some of the results might not be fully converged. In addition, given the susceptibility of hydrodynamic simulations to subgrid modelling choices, the real distribution of baryons and feedback processes might not be captured by the simulations explored in this work.

\section{Conclusion}
\label{sec:conclusion}
In this study, we have employed the BCM originally proposed in \citet{schneider_new_2015} and extended in \citet{schneider_baryonic_2019,giri_emulation_2021} to model the host contribution to the observed FRB DM. Our extended BCM is based on eight free parameters and allows for joint predictions of the gas and stellar density profiles of a halo of given mass and redshift, the two key ingredients needed to model FRB host DMs. Assuming that the distribution of FRBs within an individual halo follows the stellar density, we have used our model to predict the PDF of the host DM by sampling FRBs in the halo and integrating along the line-of-sight to a distant observer. 

Comparing our model to results derived from the CAMELS suite of hydrodynamic simulations, we find it to well reproduce the host DM PDFs as obtained from both the SIMBA and IllustrisTNG realisations. In addition, the parameter sets required to fit the host DM PDFs allow us to predict the matter power spectrum suppression due to baryonic feedback obtained in both simulations within cosmic variance. This provides an important consistency test of the extended BCM presented in this work. Furthermore, we find both the mass- and redshift dependence of the mean DM observed in CAMELS to be rather well reproduced by our model. In particular, we find the mean DM to increase with halo mass, and we also reproduce the mild DM increase with lower redshift as observed in the simulations.

Our analytic approach allows us to investigate the dependence of the host DM PDF on the parameters of our model; and our results indicate it to be quite sensitive to the parameters of the BCM. In general, the shape of the host DM PDF is determined by the interplay between the stellar and the gas distribution in halos. The degeneracy between these two components means that the features observed in hydrodynamic simulations can be reproduced in two different ways: Steep and compact FRB distributions in halos require a shallower gas density profile, while on the other hand a shallower distribution of the FRB population requires gas densities more concentrated in the halo centre. In the parlance of baryonic effects, the former corresponds to comparatively stronger feedback and the latter to weaker
feedback. However, using the DM PDF by itself, we cannot distinguish between these two cases without strong priors on the FRB distribution. However, strong feedback cannot be compensated by ever more compact stellar distributions, and the same applies in reverse for weak feedback. Therefore, the degeneracy between both profiles has its limits.
 
Limitations of the current modelling include the simplistic stellar profile. Additionally, while the model demonstrates success in fitting observables across specific mass ranges, extending this to a broader mass spectrum could necessitate a more complex approach. The omission of cold gas components in central halos is another limitation, as these could influence DM despite their typically smaller contribution due to being partially neutral. Furthermore, the accuracy of the hydrodynamic simulations used for comparison is uncertain in its own right. 

In conclusion, our findings suggest that the host galaxy's contribution to the DM of FRBs can serve as a powerful tool to explore various astrophysical processes. Specifically, this contribution could provide insights into feedback mechanisms within galaxies and the complex interactions between the circumgalactic medium and the intergalactic medium by jointly fitting stellar and gas profiles. Additionally, it could also probe the relationship between halo gas and the stellar components of galaxies. Furthermore, our model permits self-consistent predictions of the host DM PDF and the suppression of the matter power spectrum due to baryonic effects, making it promising for modelling host-DM-related systematics in FRB analyses. 

The current model, while effective in predicting the DM contribution from host galaxies, focuses primarily on the properties within individual halos. Our future objective is to improve this model to account for the entire line-of-sight, incorporating contributions from all intervening structures between the FRB source and the observer. This approach aims to develop a more consistent and accurate model for the DM PDF of FRBs.

\section*{Acknowledgements}
\review{
The authors would like to thank an anonymous referee for carefully reading the manuscript and helping to improve it significantly.}
SH was supported by the Excellence Cluster ORIGINS which is funded by the Deutsche Forschungsgemeinschaft (DFG, German Research Foundation) under Germany’s Excellence Strategy - EXC-2094 - 390783311.

\bibliographystyle{mnras}
\bibliography{MyLibrary,more_bib} 

\renewcommand\thesection{\alpha{section}}
\renewcommand\thesubsection{\thesection.\arabic{subsection}}
\renewcommand\thefigure{\thesection.\arabic{figure}} 
\setcounter{figure}{0}
\newpage
\newpage
\appendix

\section{The two-halo term}
\label{app:two_halo}
\begin{figure}
\vspace{.5cm}
    \centering
    \includegraphics[width = 0.45\textwidth]{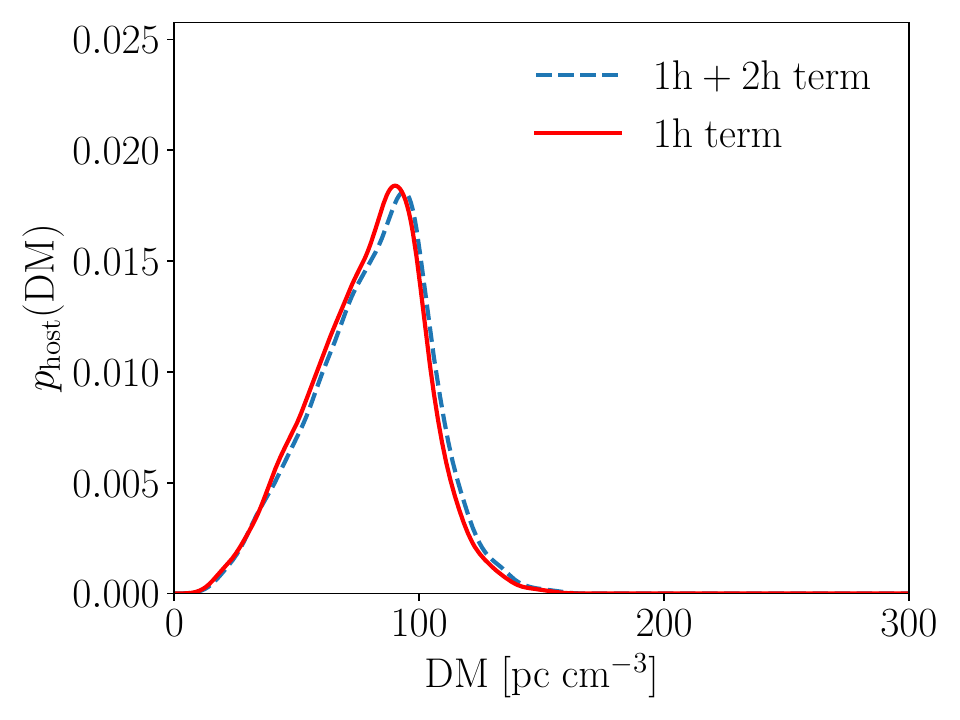}
    \caption{Comparison of the host DM PDF for a halo of viral mass $m_\mathrm{vir} = 10^{12}\;h^{-1}M_\odot$ and at redshift $z=0$ for the fiducial parameters of the BCM, including only the one-halo term in the electron distribution (red solid), and including both the one-halo and the two-halo term (blue dashed).}
    \label{fig:pdf_dm_2h}
\end{figure}

\Cref{fig:pdf_dm_2h} demonstrates that the effect is completely negligible, as the one-halo term matches the total profile of one- and two-halo term. This is likely due to the sampling distribution of FRBs, which are primarily sampled within in the core of the halo and therefore always obtain the most of their contribution from the dominant one-halo term. 

\end{document}